\documentclass[fleqn,usenatbib]{mnras}

\usepackage{newtxtext,newtxmath}

\newcommand\cag{$^{12}{\rm C}(\alpha,\gamma)^{16}{\rm O}\,$~}
\newcommand\nickel{$^{56}\mathrm{Ni}$~}

\usepackage[T1]{fontenc}
\usepackage{subfigure}
\usepackage[normalem]{ulem}
\usepackage[rightcaption]{sidecap}

\usepackage{supertabular}
\usepackage{dcolumn}

\DeclareRobustCommand{\VAN}[3]{#2}
\let\VANthebibliography\thebibliography
\def\thebibliography{\DeclareRobustCommand{\VAN}[3]{##3}\VANthebibliography}

\usepackage{graphicx}	
\usepackage{amsmath}	

\title[\cag rate on \nickel synthesis in PISNe]{Impacts of the \cag reaction rate on \nickel nucleosynthesis in pair-instability supernovae}

\author[H. Kawashimo et al.]{
Hiroki Kawashimo,$^{1,2}$\thanks{E-mail: h-kawashimo@g.ecc.u-tokyo.ac.jp}
Ryo Sawada,$^{1,3}$
Yudai Suwa$^{1,4}$
Takashi J. Moriya,$^{5,6,7}$
Ataru Tanikawa$^{1,8}$\newauthor{}
and Nozomu Tominaga$^{5,6,9}$
\\
$^{1}$Department of Earth Science and Astronomy, Graduate School of Arts and Sciences, The University of Tokyo, Meguro, Tokyo 153-8902, Japan\\
$^{2}$RIKEN Nishina Center for Accelerator-based Science, RIKEN, Wako, Saitama 351-0198, Japan\\
$^{3}$Institute for Cosmic Ray Research, The University of Tokyo, Kashiwa, Chiba 277-8582, Japan\\
$^{4}$Center for Gravitational Physics and Quantum Information, Yukawa Institute for Theoretical Physics, Kyoto University, Kyoto 606-8502, Japan\\
$^{5}$National Astronomical Observatory of Japan, National Institutes of Natural Sciences, Mitaka, Tokyo 181-8588, Japan\\
$^{6}$Astronomical Science Program, Graduate Institute for Advanced Studies, SOKENDAI, Mitaka, Tokyo 181-8588, Japan\\
$^{7}$School of Physics and Astronomy, Faculty of Science, Monash University, Clayton, Victoria 3800, Australia\\
$^{8}$Center for Information Science, Fukui Prefectural University, Eiheiji, Fukui 910-1195, Japan\\
$^{9}$Department of Physics, Faculty of Science and
Engineering, Konan University, Kobe, Hyogo 658-8501, Japan\\
}

\date{Accepted XXX. Received YYY; in original form ZZZ}

\pubyear{2024}

\begin{document}
\label{firstpage}
\pagerange{\pageref{firstpage}--\pageref{lastpage}}
\maketitle

\begin{abstract}
Nuclear reactions are key to our understanding of stellar evolution, particularly the \cag rate, which is known to significantly influence the lower and upper ends of the black hole (BH) mass distribution due to pair-instability supernovae (PISNe). However, these reaction rates have not been sufficiently determined. We use the {\tt MESA} stellar evolution code to explore the impact of uncertainty in the \cag rate on PISN explosions, focusing on nucleosynthesis and explosion energy by considering the high resolution of the initial mass.
Our findings show that the mass of synthesized radioactive nickel (\nickel) and the explosion energy increase with \cag rate for the same initial mass, except in the high-mass edge region. With a high (about twice the {\tt STARLIB} standard value) rate, the maximum amount of nickel produced falls below 70 $M_\odot$, while with a low rate (about half of the standard value) it increases up to 83.9 $M_\odot$. These results highlight that carbon "preheating" plays a crucial role in PISNe by determining core concentration when a star initiates expansion.
Our results also suggest that the onset of the expansion, which means the end of compression, competes with collapse caused by helium photodisintegration, and the maximum mass that can lead to an explosion depends on the \cag reaction rate.
\end{abstract}




\section{Introduction} \label{sec:intro}
Pair Instability Supernovae (PISNe) are the explosive deaths of very massive stars, which have been theoretically predicted \citep[e.g.,][]{1967PhRvL..18..379B,2001ApJ...550..372F,2003ApJ...591..288H} and a good candidate has recently been discovered \citep{2024A&A...683A.223S}.
In very massive stars that form massive helium cores \citep[$M_\mathrm{He}\gtrsim45M_\odot$;][]{2002ApJ...567..532H}, the electron-positron creation reactions take place in the core soften the equation of state, and reduce the adiabatic index $\gamma$ below $4/3$ \citep{1968Ap&SS...2...96F}. To be specific, thermal energy is converted into the rest mass of the electron-positron pairs, decreasing the pressure \citep[][]{1967ApJ...148..803R}. The instability induced by this pressure reduction causes the core to collapse, leading to explosive oxygen and silicon burning \citep[][]{1967ApJ...150..131R}. If the explosive oxygen burning provides enough energy, its thermonuclear energy can reverse the collapse, leading the entire star to explode with no remnant behind it. 
It is also predicted from stellar evolutionary theory that when massive progenitors become PISNe, we can observe the luminous transients ($10^{44}$ erg s$^{-1}$ or brighter at peak) for several months \citep[e.g.,][]{2002ApJ...567..532H,2005ApJ...633.1031S,2011ApJ...734..102K,2013MNRAS.428.3227D}.

Since a PISN completely destroys stars and leaves no compact objects behind, it has been thought that there is a pair-instability mass gap in the black hole mass distribution at $50-130M_{\odot}$, corresponding to the progenitors of the mass region where PISN occurs \citep{2002ApJ...567..532H,2007Natur.450..390W,2016A&A...594A..97B,2017ApJ...836..244W,2019ApJ...878...49W,2017MNRAS.470.4739S}. Hence, the upper limit of the mass gap is considered to be determined by the mass range of PISNe and the lower limit by the transition between PISNe and pulsational pair-instability supernovae (PPISN) \citep[cf.][]{2020ApJ...902L..36F}. However, this conjecture is now challenged by GW190521 which has two black holes with masses of $66^{+17}_{-18}M_{\odot}$ and $85^{+21}_{-14}M_{\odot}$
\citep{2020PhRvL.125j1102A,2020ApJ...900L..13A,2022ApJ...924...79E}, and the PISN condition is required to be reconsidered \citep[cf.][]{2021ApJ...907L...9N,2024PhRvD.109b2001A,2021MNRAS.501L..49K,2023MNRAS.522.1686M}.

The \cag reaction rate is one of the most influential nuclear reactions in the evolution of stars \citep{2009ApJ...702.1068T,2010ApJ...718..357T}, and this is also true for PISNe \citep{2018ApJ...863..153T}. However, the \cag reaction rate is difficult to determine experimentally with the current measurement sensitivity and remains highly uncertain \citep{2017RvMP...89c5007D}. Therefore, it is important to perform astrophysical simulations that take this uncertainty into account \citep[e.g.,][]{1993PhR...227...65W,2015PTEP.2015f3E01K,2022ApJ...924...39M,2022ApJ...937..112F}. 

Recently, the uncertainty in the \cag reaction rate was found to affect the range of PI mass gaps \citep{2019ApJ...887...53F,2020ApJ...902L..36F,2021MNRAS.501.4514C} \citep[cf.][]{2022ApJ...924...39M}. It suggested that black holes can be generated in mass regions previously thought to be PI mass gaps, and has attracted attention in explaining GW190521\footnote{Note that there are many suggestions to fill the PI mass gaps without changing \cag reaction rate \citep[e.g.][]{2019PhRvD.100d3027R, 2020MNRAS.497.1043D, 2020ApJ...904L..26F,2020ApJ...905L..21U,2021ApJ...908L..29G,2021PhRvL.126e1101D,2021JCAP...07..032C, 2021MNRAS.505.2170T,2021PhRvD.104d3015Z,2022MNRAS.512..884R,2022MNRAS.516.1072C,2022ApJ...941..100S,2022arXiv221213903Z,2023MNRAS.522.1686M,2023ApJ...944...40V}.}. 
From there, when considering stellar mass distribution, it is expected that the \cag reaction rate also affects the event rate of PISNe \citep{2023MNRAS.519L..32T}. Thus, the effect of the \cag reaction rate on PISNe is a noteworthy issue from the standpoint of optical observations. However, it is not clear how the uncertainties of the \cag reaction rate affect the brightness of individual PISNe.

The amount of radioactive nickel \nickel that determines the brightness of an SN is important as information is directly related to observations. It will be helpful to predict the detectability of PISNe by upcoming observatories \citep{2019PASJ...71...59M,2020ApJ...894...94R,2022A&A...666A.157M,2022ApJ...925..211M, 2023MNRAS.519L..32T,2023MNRAS.520..866A}. In addition, nickel synthesis is also an important topic from galactic chemical evolution since nickel is eventually turned into iron and supplied to space. In this study, we have used stellar evolution calculations to consider PISNe that occur under various \cag rates and calculate the amount of \nickel produced and the explosion energy.

This paper is structured as follows. In section \ref{sec:mam}, we explain the investigation methods. In section \ref{sec:results}, we show our results and discuss our findings. We conclude the paper in Section \ref{sec:summary}.

\section{Models and Methods} \label{sec:mam}
\subsection{Setup}
We utilize version 15140 of the stellar evolution code {\tt MESA} \citep{2011ApJS..192....3P,2013ApJS..208....4P,2015ApJS..220...15P,2018ApJS..234...34P,2019ApJS..243...10P,2023ApJS..265...15J} to simulate the evolutionary process of helium cores. These cores either collapse to form black holes or undergo explosive events known as Pair-Instability Supernovae (PISNe). The input parameter configuration is based on the default model choices outlined by \cite{2019ApJ...882...36M}, specifically referred to as the {\tt ppisn} setup within {\tt MESA-r15140} \footnote{We note that one alteration from the original {\tt ppisn} setup involves omitting inlist switching based on helium depletion to avoid potential failures during the handoff between inlists.}. Note that we determined the success or failure of PISN using the same criteria as in \cite{2019ApJ...882...36M}.
We suppose that a PISN succeeds when all parts of the star exceed the escape velocity, and the calculation is terminated at that time. We also determine failure based on the central density exceeding $10^{12}\;\mathrm{g\;cm^{-3}}$ and the maximum infall velocity of the central Fe core exceeding $8\times 10^{8}\;\mathrm{cm\;s^{-1}}$.

In our simulations, we initiate the process by employing a non-rotating model of hydrogen-free helium stars with a metallicity of $Z=10^{-5}$. 
Given our specific focus on understanding the \nickel amount and explosion energy in the PISN explosions and resolving the transition between successful PISN and CC models, we conducted calculations using various initial mass ranges. We initially explored a broad range of initial masses, spanning from 40 to 180 $M_\odot$, with increments of 5 $M_\odot$. Within this range, the occurrence of PISN explosions was confirmed through calculations performed in increments of 1 $M_\odot$. Furthermore, we conducted simulations with finer resolution, using increments of 0.1 $M_\odot$ near the upper boundary of the mass range and subsequently employing increments of 0.01 $M_\odot$ in the immediate vicinity of the uppermost edge (see Appendix \ref{sec:appendix-table}). Our investigated mass range covers between 70 and 150 $M_\odot$ near the region of the PISN BH mass gap, as revealed by previous studies \citep{2019ApJ...882...36M}.

The evolution of helium stars serves as a valuable laboratory for investigating the evolution of massive stars experiencing pair-instability. This is because a majority of massive stars are believed to have shed their outer hydrogen layers, thereby exposing their helium cores. Furthermore, the properties of these stars in their final phase are strongly influenced by the mass of their helium cores \citep{2017ApJ...836..244W,2019ApJ...882...36M}. It is important to note that progenitors of merging binary black holes also undergo the loss of their hydrogen envelopes as a result of binary interactions unless their metallicity is nearly zero or convective overshoot is ineffective \citep[e.g.,][]{2022ApJ...926...83T}. 

We utilize the {\tt approx21\_plus\_co56.net} nuclear reactions network integrated into the {\tt MESA} framework. This network has been proven to be efficient and accurate in estimating explosion energy and the quantity of synthesized \nickel during explosive nucleosynthesis \citep{2010NuPhA.841....1L,2013ApJS..207...18S,2015JPhG...42c4007I,2016ApJ...831..107I,2019ApJ...887...53F}.
For nuclear reaction rates, we adopt the default rates provided by {\tt MESA} in this version, which are based on {\tt NACRE} \citep{1999NuPhA.656....3A} and {\tt JINA REACLIB} \citep{2010ApJS..189..240C}. However, there is one exception, namely the \cag rate, which is discussed in detail in Section \ref{sec:mam-cag}.

For hydrodynamics, the setup uses the HLLC method, which is useful for modeling shock waves \citep{1994ShWav...4...25T}. The simulation is switched from hydrostatic to dynamical when the stellar global stability index falls below its critical value, $4/3$. This index is calculated using the local pressure $P$ and the local density $\rho$, as represented by the equation below:
\begin{align}
\langle\Gamma_1\rangle=\dfrac{\int_0^M\frac{\Gamma_1 P}{\rho}dm}{\int_0^M\frac{P}{\rho}dm},
\label{eq:faea}
\end{align}
where $\Gamma_1$ is the local first adiabatic exponent. This corresponds to the time when neutrino cooling is progressing rapidly \citep[cf.][]{2019ApJ...882...36M,2019ApJ...887...53F}.

\subsection{The treatment of the \cag rate}\label{sec:mam-cag}

The treatment of the \cag rate is the most important part of this paper, and it is essentially based on the previous studies by \citet{2019ApJ...887...53F,2020ApJ...902L..36F}.
We utilize {\tt STARLIB} reaction rate library, which provides the median nuclear reaction rate, $\langle\sigma_\mathrm{c.s.} v\rangle_\mathrm{med}$, and the associated uncertainty factor, ${\tt f.u.}$, at temperatures ranging from $T=10^6$ to $10^{10}$ K \citep{2013ApJS..207...18S}.  Following the approach of \citet{2010NuPhA.841....1L},  we assume that all reaction rates provided by {\tt STARLIB} follow a log-normal probability distribution. The log-normal distribution is characterized by the position parameter $\mu$ and spread parameter $\sigma$, respectively.
\begin{equation}
    P(x)=\cfrac{1}{\sqrt{2\pi\sigma^2}x}\exp{\left(-\cfrac{(\ln{x}-\mu)^2}{2\sigma^2}\right)} ~.
\end{equation}
These parameters can be obtained using the median rate $\langle\sigma_\mathrm{c.s.} v\rangle_\mathrm{med}$ and the factor uncertainty {\tt f.u.} represented in {\tt STARLIB} as follows.
\begin{align}
\mu &= \ln{(\langle\sigma_\mathrm{c.s.} v\rangle_\mathrm{med})} ~,\\
\sigma &= \ln{\tt (f.u.)} ~.
\end{align}
In a lognormal distribution, the natural logarithm of the random variable ($y=\ln{x}$) follows a normal distribution. The parameters $\mu$ and $\sigma$ represent the mean and standard deviation of the corresponding normal distribution, respectively. Therefore, in this context, we parameterize the \cag reaction in terms of the number of sigmas, $\pm n\cdot\sigma$, from the median {\tt STARLIB} \cag reaction rate:
\begin{align}
   \langle\sigma_\mathrm{c.s.} v\rangle_{n\cdot\sigma}&\equiv\exp{\left( \mu+n\cdot\sigma\right)} \nonumber \\
    &=\langle\sigma_\mathrm{c.s.} v\rangle_\mathrm{med}\cdot({\tt f.u.})^n ~.
\end{align}
\begin{figure}
    \centering
    \includegraphics[width=0.48\textwidth]{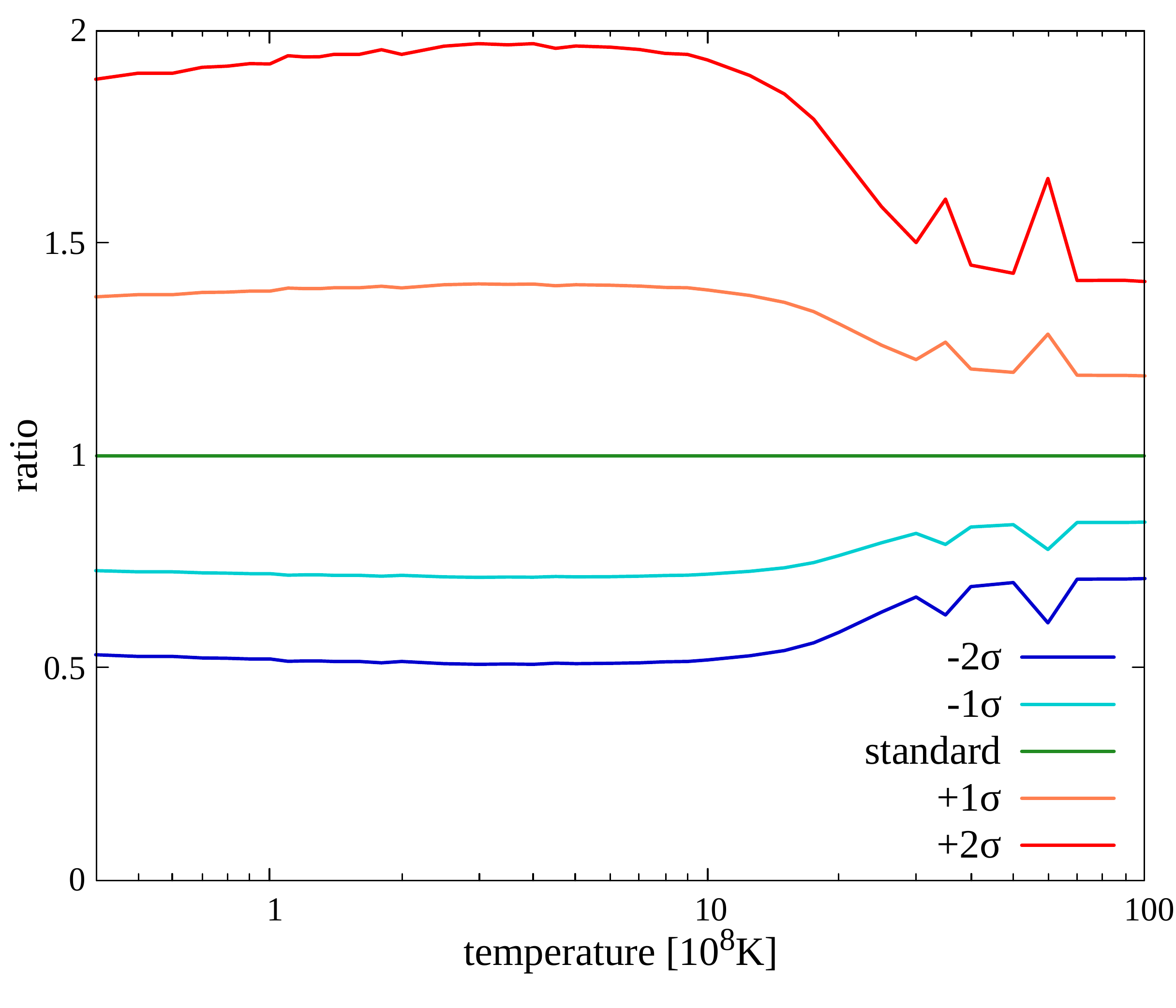}
    \caption{
    The \cag rate as a function of temperature, normalized to the median rate $\langle\sigma_\mathrm{c.s.} v\rangle_{\pm n\cdot\sigma}/\langle\sigma_\mathrm{c.s.} v\rangle_\mathrm{med}$ from {\tt STARLIB}. $\langle\sigma_\mathrm{c.s.} v\rangle_\mathrm{med}$ and its uncertainty are from \citet{2002ApJ...567..643K}. The color convention for the \cag rate remains consistent throughout our paper.}
    
    \label{fig:ratio}
\end{figure}

Figure \ref{fig:ratio} shows the \cag rate as a function of temperature, normalized to the median {\tt STARLIB} rate $\langle\sigma_\mathrm{c.s.} v\rangle_{\pm n\cdot\sigma}/\langle\sigma_\mathrm{c.s.} v\rangle_\mathrm{med}$.
$\langle\sigma_\mathrm{c.s.} v\rangle_\mathrm{med}$ and its uncertainty are from \cite{2002ApJ...567..643K}.
Hereafter, when referring to the reaction rate $\langle\sigma_\mathrm{c.s.} v\rangle_{\pm n\cdot\sigma}$, we simply denote it a $\pm n\cdot\sigma$. To examine the effects of  \cag burning rate, we simulate stellar models using calculated \cag rates ranging from $-2\sigma$ to $+2\sigma$ in increments of $1\sigma$. It is important to note that we refer to the $0\sigma$ series --- representing the most probable values --- as the {\it standard} series.

\section{Results} \label{sec:results}
\subsection{Overviews for PISN}\label{sec:results-overview}

\begin{figure}
    \centering
    \includegraphics[width=0.48\textwidth]{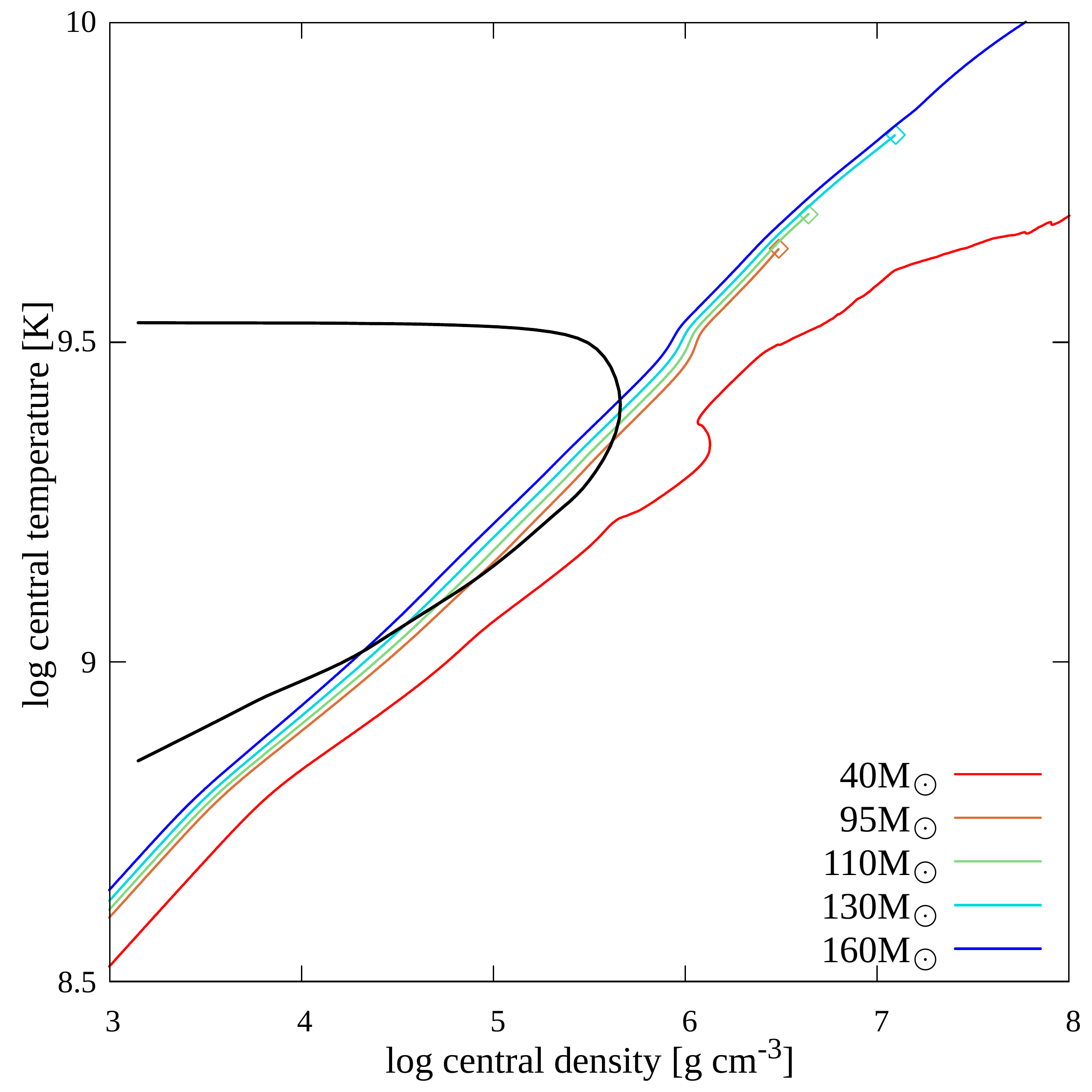}
    \caption{$\rho_\mathrm{c}$-$T_\mathrm{c}$ trajectories for models with initial He core masses of $40$, $95$, $110$, $130$, and $160M_{\odot}$, using the standard \cag rate. Each color corresponds to the initial mass of the progenitor, except for the black line, which represents $\gamma=4/3$, the border of gravitational instability. The 40 and $160\, M_\odot$ models undergo core collapse, while the other models result in PISN explosions. Square points indicate the maximum temperature experienced in exploding models, marking the beginning of the expansion. Beyond these points, the trajectories of explodable models turn back adiabatically.} 
    \label{fig:rho-t}
\end{figure}

In this section, we begin by discussing the typical characteristics of PISNe and the reliability of our explosion model using the standard \cag rate.
Figure \ref{fig:rho-t} presents the central density and temperature $(\rho_\mathrm{c}$-$T_\mathrm{c})$ trajectories of various stars with different initial He core masses. The $M_{\rm init,He}=40\,M_\odot$ model and the $160\,M_\odot$ model both experience iron-core collapse, whereas the other models resulted in PISN explosions. From the figure, it is evident that the $\rho_\mathrm{c}$-$T_\mathrm{c}$ of models exceeding $95\,M_\odot$ enter into the $\gamma<4/3$ region, whereas the $40\,M_\odot$ model does not.\footnote{The global stability of a star is determined by the averaged first adiabatic exponent, see Equation \ref{eq:faea}.}

\subsection{Effects of the \cag reaction rate uncertainty}\label{sec:results-main}

In Section \ref{sec:results-main}, we provide the findings regarding the correlation between the \cag rate and the properties of the PISN explosion, specifically the explosion energy, as final total energy (Section \ref{sec:expl}) and the synthesis of nickel (Section \ref{sec:ni}). Subsequently, we explore the underlying physics behind these correlations in Section \ref{sec:cor}. All results are presented in tabular form in Appendix \ref{sec:appendix-table}. We note that the total energy is the sum of kinetic, gravitational, and internal energy. In the final phase, the stars are sufficiently expanded, and no gravitational binding so that the explosion energy is approximately equal to the kinetic energy.

\subsubsection{Explosion energy}\label{sec:expl}
\begin{figure}
    \centering
    \includegraphics[width=0.48\textwidth]{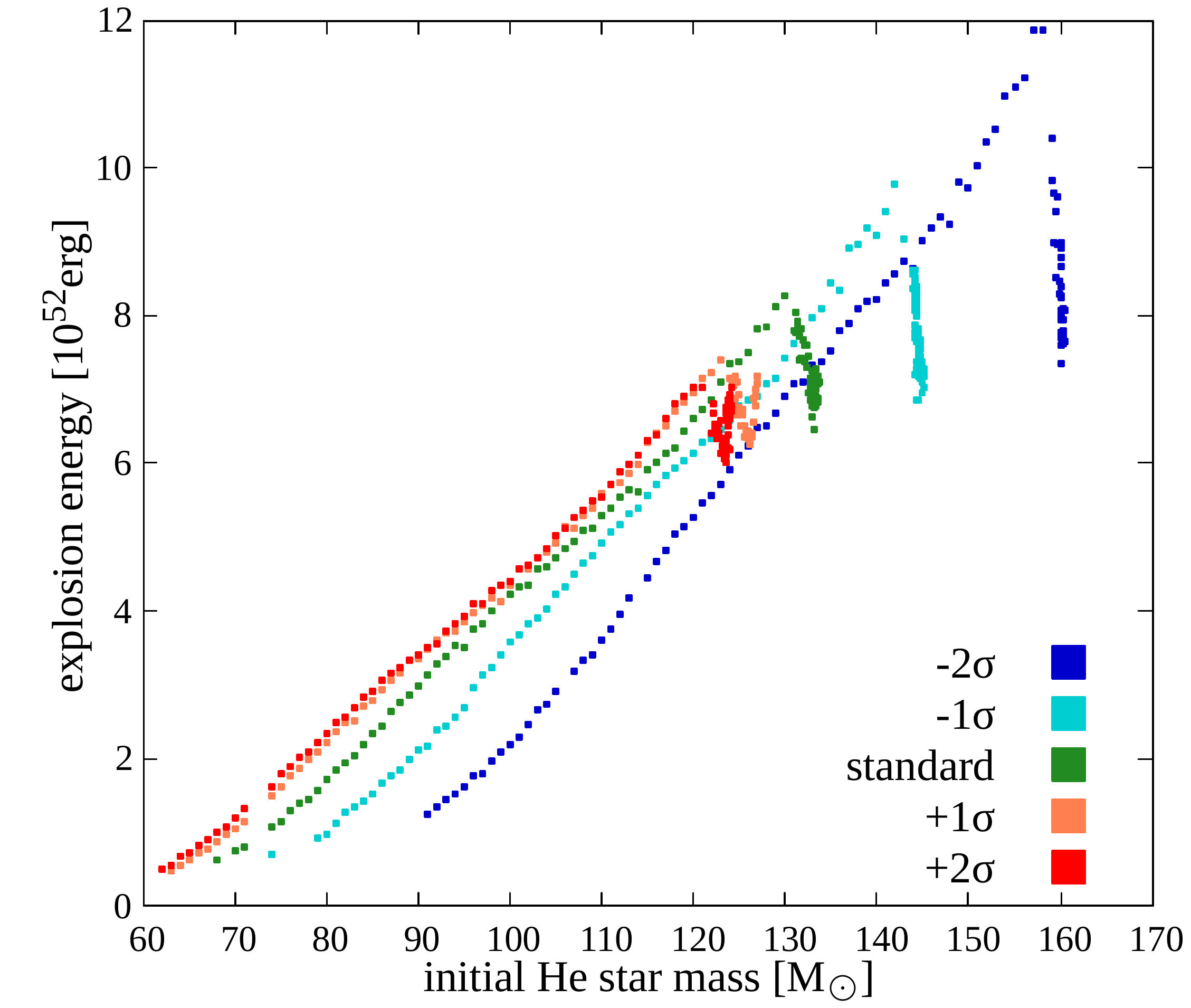}
    \caption{
    The relationship between the explosion energy $E_\mathrm{expl}$ and the initial He core mass $M_\mathrm{init,He}$ for different \cag reaction rates. Each color in the plot corresponds to a specific reaction rate as described in Figure \ref{fig:ratio}.} 
    \label{fig:mass-energy}
\end{figure}

Figure \ref{fig:mass-energy} illustrates the relationship between the explosion energy $E_\mathrm{expl}$ and the initial He core mass $M_\mathrm{init,He}$ for each \cag reaction rate. Each color corresponds to a different reaction rate. When we fix the initial He core mass, we observe that models with higher \cag rates exhibit higher explosion energies. Furthermore, within each series of the same \cag rate, we observe a consistent pattern: the explosion energy gradually increases on the low-mass side and then sharply decreases in the high-mass region (for more discussion, see Appendix \ref{sec:appendix-HW02}). This behavior is observed across all models. In the increasing trend region, we also observe that the maximum explosion energy increases as the \cag rate decreases.

\subsubsection{\nickel synthesis}\label{sec:ni}

\begin{figure}
    \centering
    \includegraphics[width=0.48\textwidth]{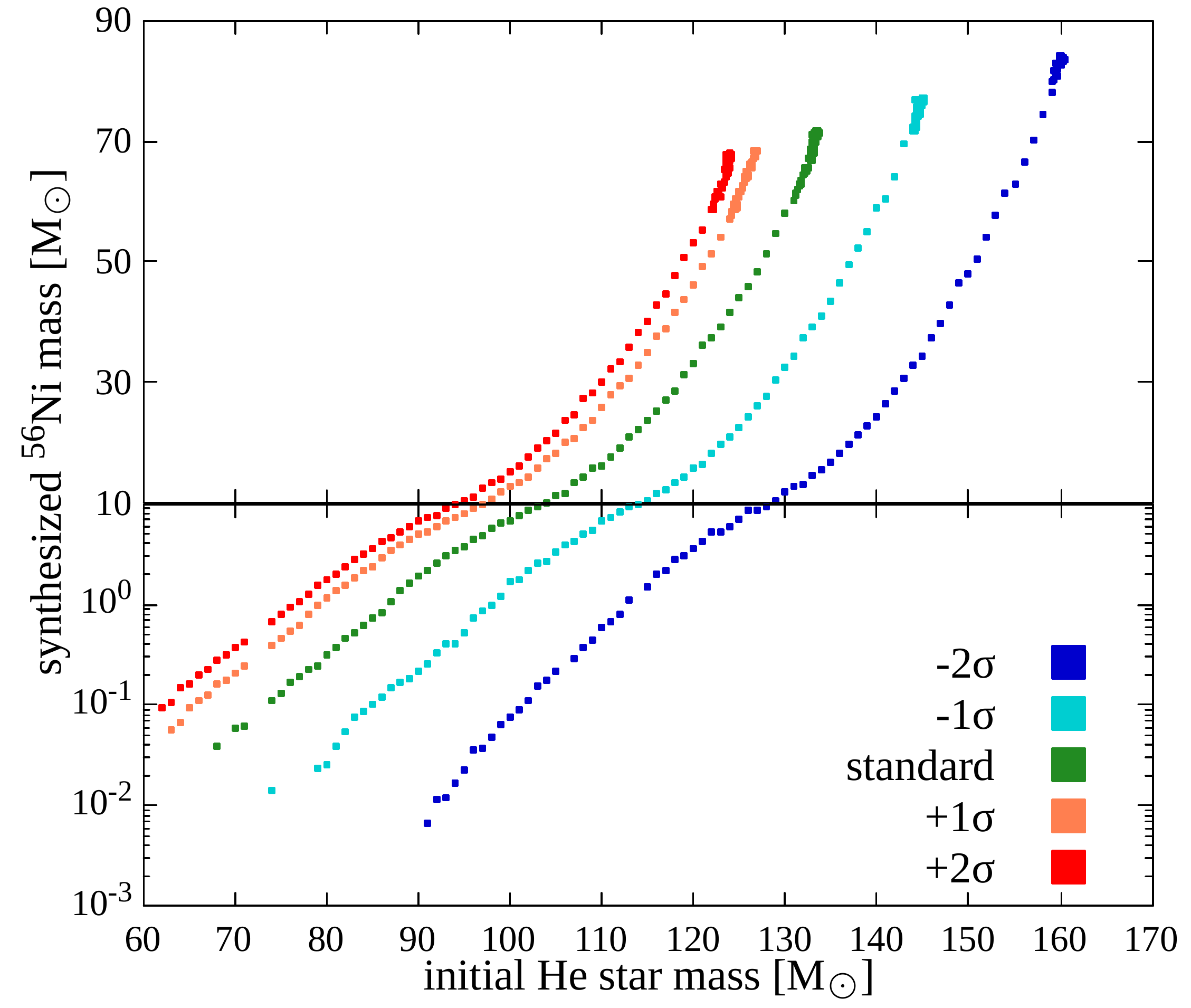}
    \caption{
    The same plot as Figure \ref{fig:mass-energy}, but for the synthesized radioactive nickel mass $M_{^{56}{\rm Ni}}$ at the final step.}
    \label{fig:nimass}
\end{figure}

Figure \ref{fig:nimass} displays the synthesized nickel mass at the final step as a function of the initial helium star mass.\footnote{Note that the nickel mass reaches its peak within approximately 100 seconds after the central temperature ($T_\mathrm{c}$) surpasses $10^{9.5}$ K. The amount of nickel remains constant until the completion of the calculation.} Notably, within the models sharing the same initial mass, a higher \cag rate results in increased nickel synthesis. Similar to the explosion energy, we observe that the amount of synthesized nickel in the most massive progenitors is greater at lower \cag reaction rates. However, we do not observe a point where the trend abruptly changes within each series.


\subsubsection{Carbon "preheating"}\label{sec:cor}

Our findings reveal that within the same progenitor mass, a higher \cag rate leads to increased total energy and the synthesis of radioactive nickel. This observation aligns with previous studies \citep{2018ApJ...863..153T,2020ApJ...902L..36F}, which suggest that these trends with the \cag rate stem from the carbon-burning process preceding the explosive oxygen burning that triggers PISNe.

\begin{figure}
    \centering
    \includegraphics[width=0.48\textwidth]{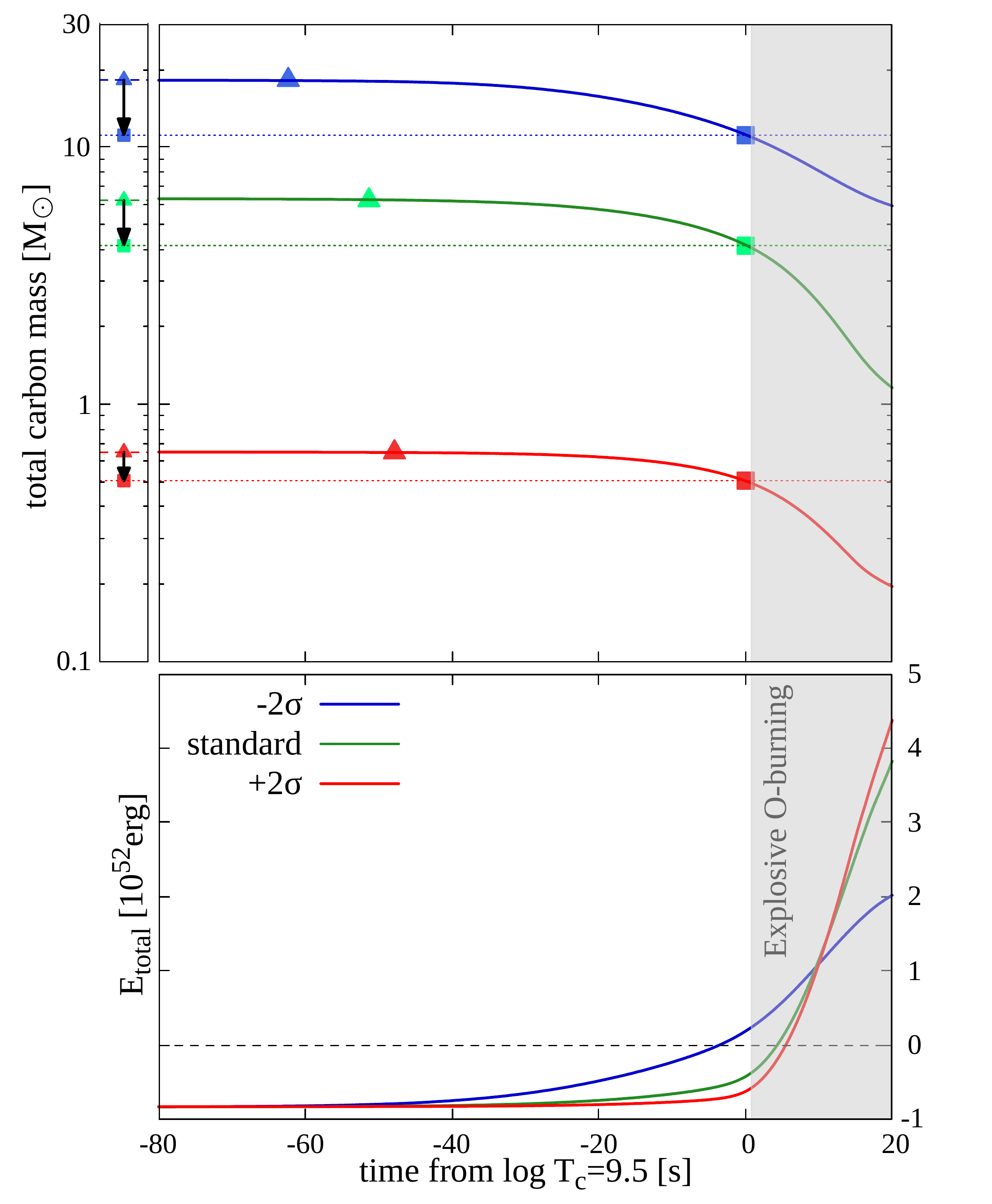}
    \caption{The time evolution of total carbon mass (top panel) and the total energy (bottom panel) for the initial He core mass $M_\mathrm{He}=100M_{\odot}$, comparing the standard and $\pm2\sigma$ \cag rate models. The time origin $t=0$ is defined as the moment when the central temperature $T_\mathrm{c}$ reaches $\log T_\mathrm{c}(\mathrm{K})=9.5$ in each model, marking the onset of explosive oxygen burning that triggers the PISN. Squares represent the residual carbon mass at $\log T_\mathrm{c}(\mathrm{K})=9.5$, while triangles correspond to $\log T_\mathrm{c}(\mathrm{K})=9.3$, which is the beginning of neon burning, it consumes residue of carbon burning. The left panel displays the difference between them. In the bottom panel, the dashed horizontal line indicates the transition between negative and positive total energy.}
    \label{fig:preheat-carbon}
\end{figure}

We describe the "preheating" process by observing energy gaining just before oxygen burning. Figure \ref{fig:preheat-carbon} presents the time trajectories of the total carbon mass and total energy for the initial He core mass $M_\mathrm{init,He}=100M_{\odot}$, in comparison to the standard \cag rate and $\pm2\sigma$ models. The time $t=0$ corresponds to when the central temperature $T_\mathrm{c}$ reaches $\log T_\mathrm{c}(\mathrm{K})=9.5$ in each model, marking the onset of explosive oxygen burning \citep{1970ApJ...160..181T,1973ApJS...26..231W}.
At high \cag reaction rates $(+2\sigma)$, the carbon is already depleted at the end of helium burning $(t \approx -80 \mathrm{s})$. Consequently, limited carbon burning occurs, and the energy remains stagnant until the onset of explosive oxygen burning. In contrast, at low \cag rates, a substantial amount of carbon persists, leading to carbon preheating that boosts the total energy prior to explosive oxygen burning. As a result, the star becomes unbound without awaiting explosive oxygen burning, leading to a gradual growth in total energy. Note that this preheating process is considered to occur within the CO core. This discussion is consistent with the known fact that PISNe are driven by explosive oxygen burning initiated in the CO core.

\subsection{The maximum mass limit of the explosion}\label{sec:max}
\begin{figure}
	\subfigure[The entire region]{
		\includegraphics[width=0.48\textwidth]{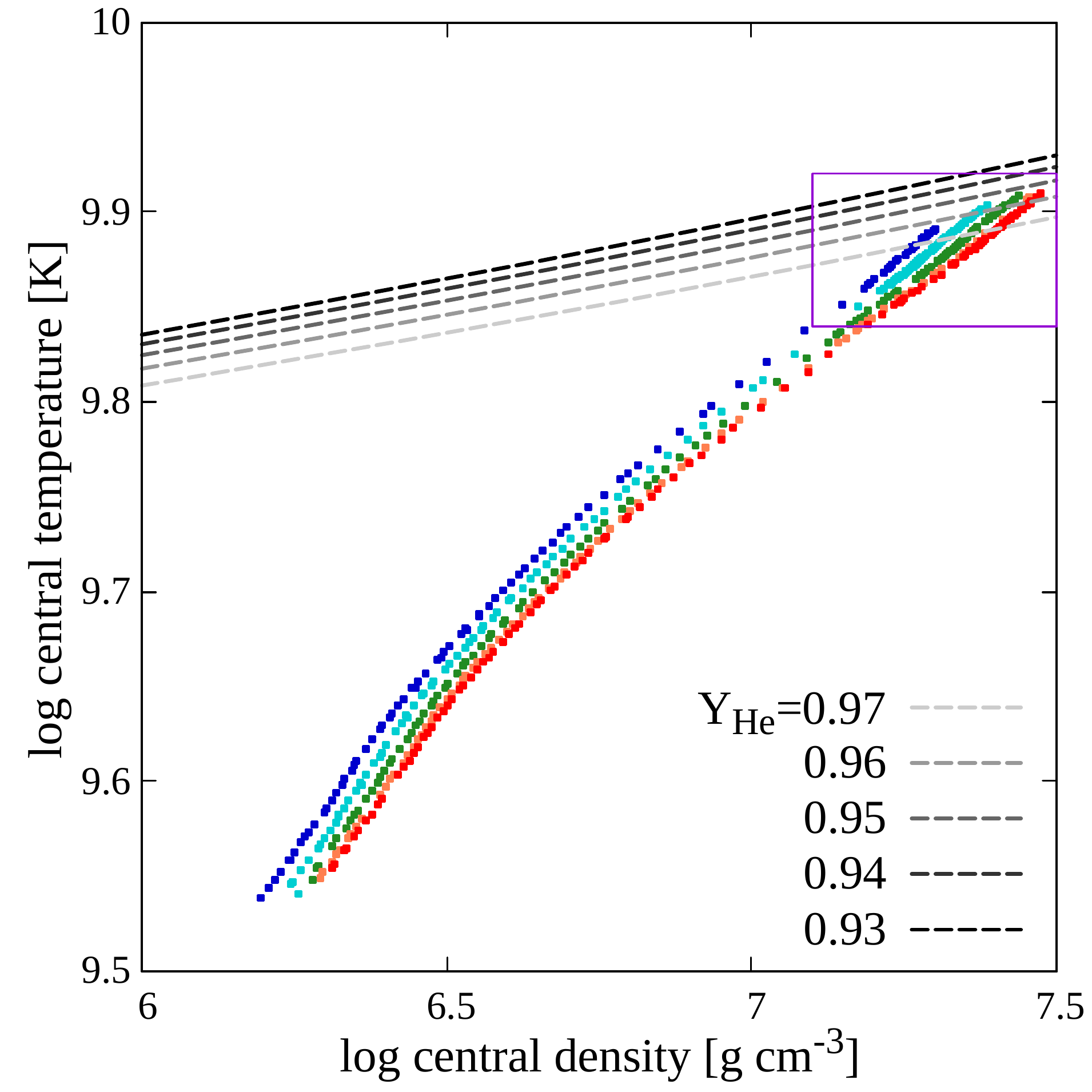}}\\ 
	\subfigure[High $\rho-T$ region]{
		\includegraphics[width=0.48\textwidth]{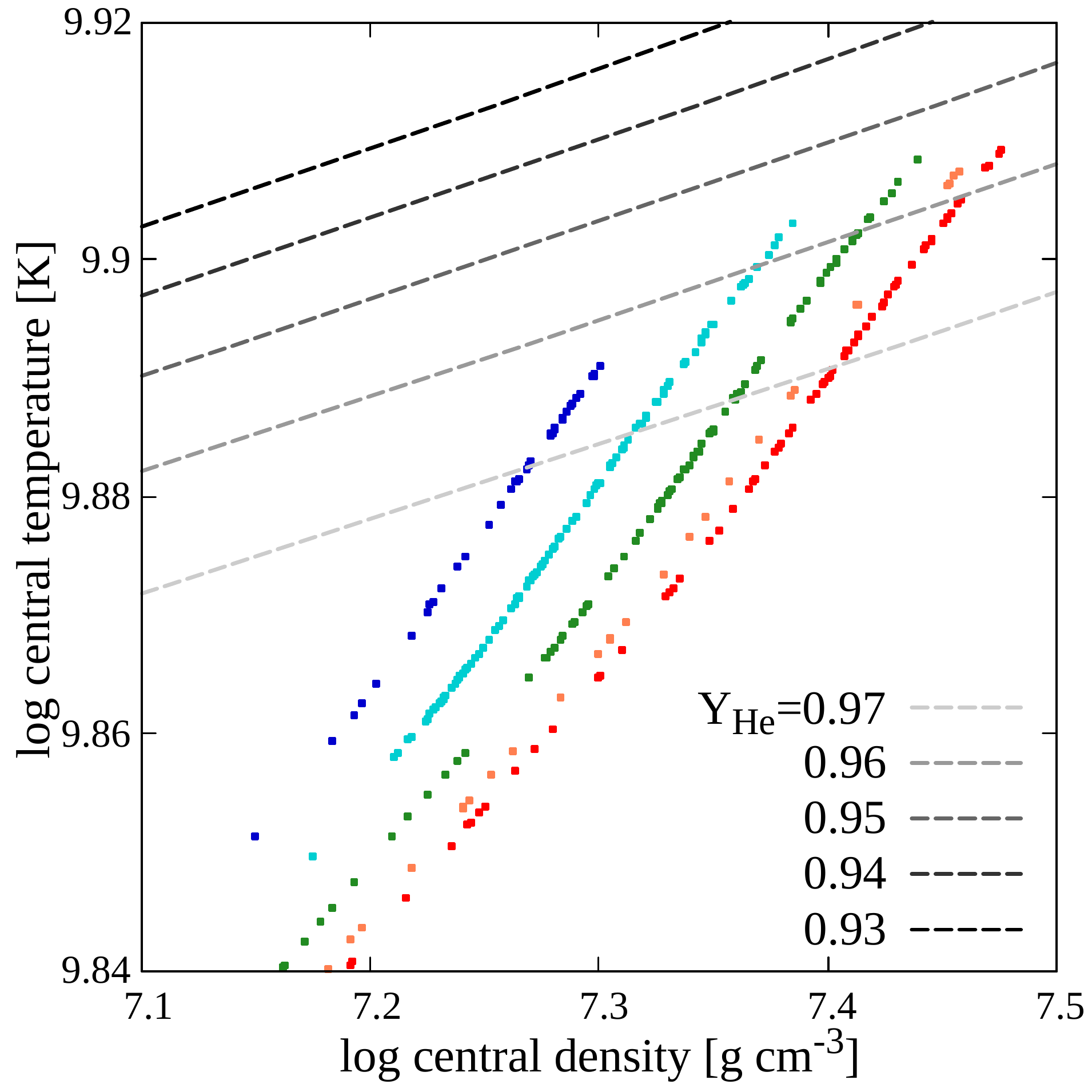}}
	\caption{The maximum central temperature and corresponding central density reached by each model. The colors represent different reaction rates, following the same convention as Figure \ref{fig:ratio}. The top panel (a) displays all exploding models, while the bottom panel (b) zooms in on the region highlighted in purple in panel (a). The grey dashed lines indicate the threshold for $^4{\rm He} \to 2{\rm n}+2{\rm p}$ photodisintegration with various $Y_{\rm He}$.}
	\label{fig:limit}
\end{figure}

In this section, we elaborate on the fact that heavier stars become explodable in low \cag rate environments. The upper limit of explodable initial mass, which represents the upper boundary of the PI mass gap, is primarily determined by photodisintegration \citep{2016MNRAS.456.1320T,2018ApJ...863..153T}. We anticipate that nickel production and decomposition will transpire concurrently within high-temperature environments. The abundance pattern of a star that undergoes a failed PISN and is just prior to collapse reveals a decrease in nickel around its center, with helium constituting the majority of the components.
The two panels in Figure \ref{fig:limit} depict the maximum central temperature and the corresponding central density experienced by each model, represented by the square points in Figure \ref{fig:rho-t}. Dashed lines in the figure represent the condition for photodisintegration of $^4{\rm He} \to 2{\rm n}+2{\rm p}$, which uses helium produced from $^{56}{\rm Ni} \to 14^4{\rm He}$.
This condition is given by
\begin{align}
    \log\left(\rho R(Y_\mathrm{He})\right)&
    =11.7974+\dfrac{3}{2}\log\left(\dfrac{k_\mathrm{B}T}{1\mathrm{MeV}}\right)
    -4.097\left(\dfrac{k_\mathrm{B}T}{1\mathrm{MeV}}\right)^{-1} \label{eq:photo}\\
    &=-3.299+\dfrac{3}{2}\log \left(\dfrac{T}{1\mathrm{K}}\right)-4.753\times10^{10}\left(\dfrac{T}{1\mathrm{K}}\right)^{-1} ~,\label{eq:photo2}
\end{align}
where 
\begin{equation}
R(Y_\mathrm{He})=Y_\mathrm{He}\left(\dfrac{1-Y_\mathrm{He}}{Y_\mathrm{He}}\right)^{4/3}~,\label{eq:defR}
\end{equation}
and $Y_\mathrm{He}$ is the residual number fraction of $^4$He (see Appendix \ref{sec:appendix-calc} for the derivation of Eq. \eqref{eq:photo} and \eqref{eq:photo2}). 
Based on Figure \ref{fig:limit}(a), it is evident that all series of \cag rates exhibit nearly identical characteristics. Furthermore, Figure \ref{fig:limit}(b) demonstrates that the transition from explosion to implosion occurs at $Y_\mathrm{He} \approx 0.96$ across all series. This finding suggest that the upper limit of PISNe is determined by the initiation of $^4{\rm He} \to 2{\rm n}+2{\rm p}$ photodisintegration.

\begin{figure}
    \centering
    \includegraphics[width=0.5\textwidth]{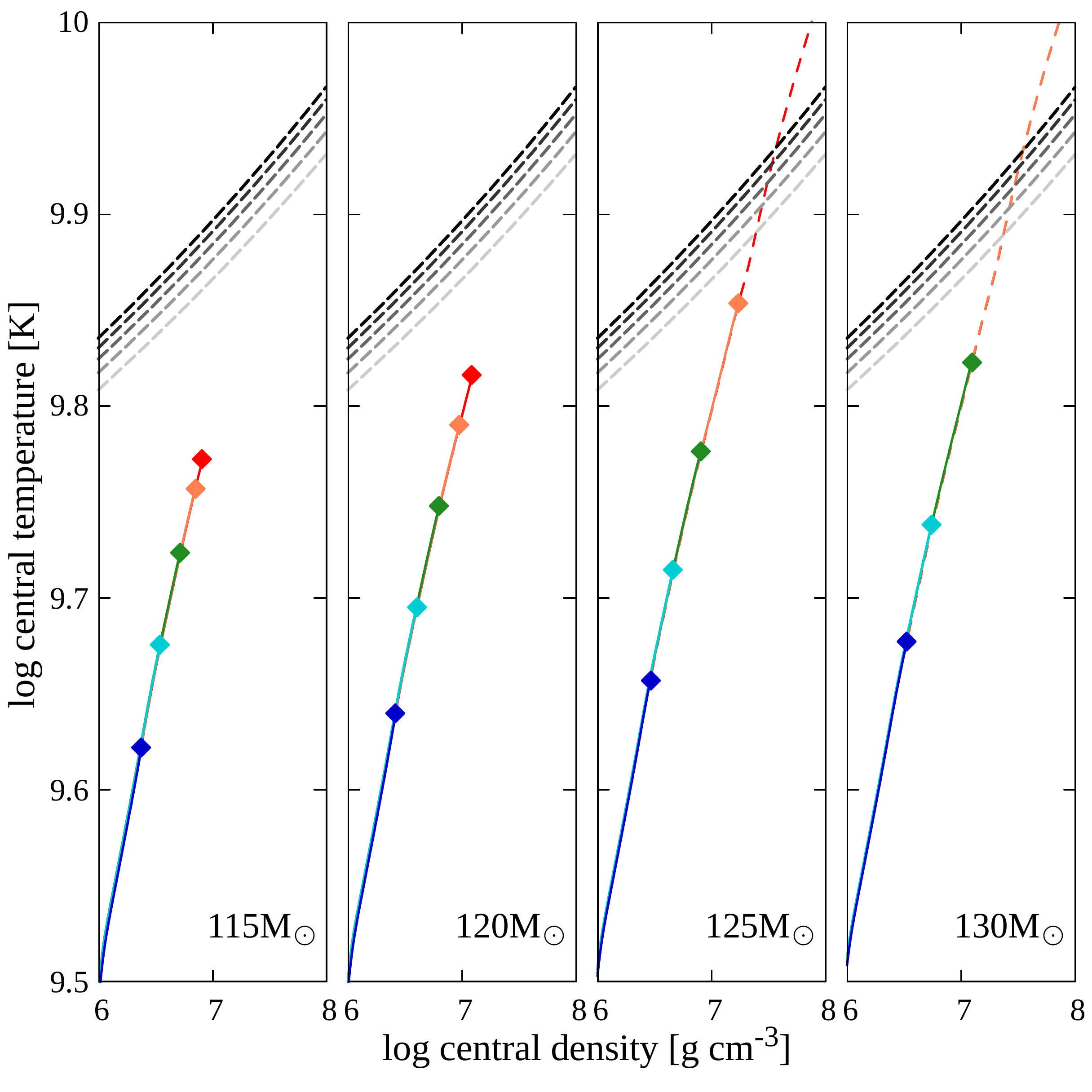}    
    \caption{The $\rho_\mathrm{c}-T_\mathrm{c}$ trajectories for different initial helium core masses: 115$M_{\odot}$ (left panel), 120$M_{\odot}$ (middle left panel), 125$M_{\odot}$ (middle right panel), and 130$M_{\odot}$ (right panel). Each trajectory is assigned a color corresponding to the reaction rate, and the grey dashed lines indicate the threshold for $^4{\rm He} \to 2{\rm n}+2{\rm p}$ photodisintegration (see Figure \ref{fig:limit}). The square points represent the endpoints, indicating the beginning of the expansion phase, it corresponds with square points in figure \ref{fig:rho-t}. The dashed lines represent the trajectories of unexploded models. It is noteworthy that in all panels, the trajectories largely overlap, as the stars undergo similar evolution regardless of the \cag rate.}
    \label{fig:multiRho-t}
\end{figure}

Figure \ref{fig:multiRho-t} shows the evolution of central density and temperature for progenitors with $M_\mathrm{He}=115,120,125,130M_{\odot}$. The square points indicate the onset of expansion (see Figure \ref{fig:rho-t}), and the trajectories of the unexploded model are depicted as dashed lines in corresponding colors.
This figure also demonstrates that the central evolution of progenitors with the same initial He mass follows a consistent trajectory regardless of the reaction rate.  The position of the onset of expansion varies. Moreover, as explained in Figure \ref{fig:rho-t}, it is evident that, for a given \cag reaction rate, the position of the onset of explosion shifts to higher temperatures and pressures as the mass increases. By combining these results with the discussion in Figure \ref{fig:limit}, it can be inferred that endpoints associated with higher \cag rates surpass the photodisintegration condition at relatively lower masses. Conversely, the endpoints for lower \cag reaction rates occur at lower temperatures and pressures, limiting only higher-mass stars to cross the photodisintegration condition. For instance, focusing on the case of 120$M_\odot$ in Figure \ref{fig:multiRho-t}, higher reaction rates undergo more intense contraction, nearing the He photodisintegration border. When considering 125$M_\odot$, any progenitors experience higher temperatures and pressures compared to the case of 120$M_\odot$, resulting in a shift in the endpoint. As a consequence, the $+2\sigma$ model exceeds the He photodisintegration border and fails to explode as PISN. Conversely, the $-2\sigma$ model, farthest from the He photodisintegration border, remains sufficiently distant even for a 130$M_\odot$ case, indicating that it would not exceed the He photodisintegration border without becoming even more massive.



\section{Summary} \label{sec:summary}

We conducted stellar evolution calculations to investigate the impact of \cag rates on \nickel nucleosynthesis in pair-instability supernovae (PISNe). Our findings indicate that lower \cag reaction rates result in a greater amount of synthesized nickel in the heaviest explodable progenitor stars. 
For instance, the upper-mass limit of the synthesized nickel mass changes from $67M_\odot$ ($+2\sigma$) to $83M_\odot$ ($-2\sigma$), corresponding to $125M_\odot$ ($+2\sigma$) and $160M_\odot$ ($-2\sigma$) for the maximum mass of exploding progenitors. The shift of those mass ranges has already found in previous studies for lower-mass side as PPISN-PISN transition line, and our findings are consistent with the same trends for these insights \citep{2020ApJ...894...94R,2021MNRAS.501.4514C,2021ApJ...912L..31W}. 
The novelty of this study lies in the systematic calculations of the synthesized nickel mass, which has not been investigated in the previous works. 
The change in the synthesized nickel mass may be attributed to the carbon preheating process. Additionally, we demonstrated that distinct \cag reaction rates give rise to varying ranges of explodable masses due to the interplay between He photodisintegration and the preheating effect.

Note that these results will be affected by the size of the nuclear reaction network. We probably overestimate the amount of nickel produced and the rate of energy absorption by photodisintegration due to the current small network \citep[see][also Appendix \ref{sec:appendix-HW02}.]{2020A&A...640A..56R,2016ApJS..227...22F}. However, this overestimation is small enough compared to the amount of change from \cag reaction varying. Our main results, therefore, are reliable even after taking into account the uncertainty of the size of the network. On the other hand, it should be noted that some previous studies have shown that larger networks synthesize more nickel \citep{2019ApJ...882...36M,2020A&A...640A..56R}.

Our findings have implications for estimating the detectability of PISNe, particularly regarding their dependence on the \cag rate \citep[e.g.,][]{2012MNRAS.422.2701P, 2019PASJ...71...59M, 2022A&A...666A.157M, 2022ApJ...925..211M, 2019PASJ...71...60W, 2020ApJ...894...94R}. \cite{2023MNRAS.519L..32T} conducted population synthesis calculations to investigate the impact of \cag rates on PISN discoveries using the Euclid space telescope \citep{2011arXiv1110.3193L}. They found that PISNe would be more frequently detected in the standard \cag case compared to the $-3\sigma$ \cag case due to a higher intrinsic PISN event rate in the former case. However, their assumptions about identical light curves for PISNe with different \cag rates raised concerns about the validity of their results.
Finally, our results can address these concerns. Figure \ref{fig:nimass} indicates that PISNe in the low \cag rate case tends to be fainter than those in the standard \cag case when the initial He star masses are fixed. Although the maximum luminosity of PISNe gradually increases as \cag rates decrease, it will not significantly impact PISN detectability. This is because PISNe with higher He star masses are already rare due to initial stellar mass functions in which the number of stars decreases with their masses increasing \citep{1955ApJ...121..161S, 2018Sci...359...69S}. In the future, we will further investigate this argument by combining binary population synthesis calculations with PISN light curves, particularly for the low \cag case.

\section*{Acknowledgments}

H. K. thanks Koh Takahashi, Kanji Mori, Tomoya Takiwaki, and Hiroki Nagakura for fruitful discussions.
This work is supported by Grant-in-Aid for Scientific Research (JP18H05437, JP20H00174, JP20H01904, JP21K13966, JP21H04997, JP22KJ0528, JP21K13964, JP22H04571) from the Ministry of Education, Culture, Sports, Science and Technology (MEXT), Japan. HK is supported by RIKEN Junior Research Associate Program.

\section*{Data availiability}
The data underlying this article will be shared on reasonable request to the corresponding author.




\appendix
\section{Additional considerations on the effect of carbon preheating on the internal structure}\label{sec:appendix-structure}

We examined the significant impact of carbon preheating prior to oxygen burning on the explosion energy in Section \ref{sec:cor}. In this appendix, we present additional findings that shed light on how carbon preheating influences the internal structure of stars. These results offer intriguing insights into the mechanisms through which carbon preheating affects both the explosion energy and the synthesis of nickel.

\begin{figure}
    \centering
    \includegraphics[width=0.48\textwidth]{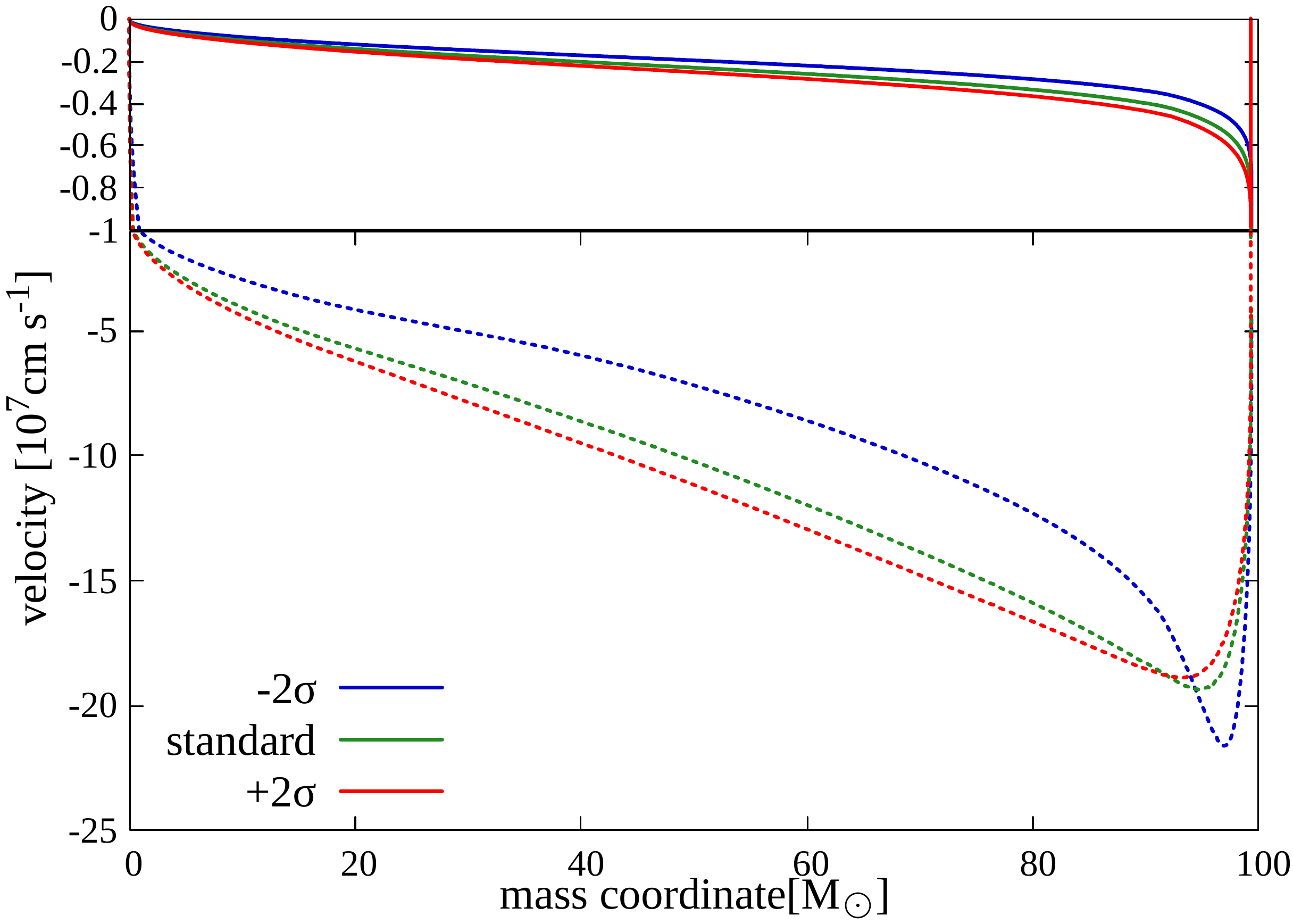}
    \caption{The velocity profiles of different parts of a 100$M_{\odot}$ model. The solid line corresponds to the velocity just before the onset of carbon preheating ($T_\mathrm{c}=9.2$), while the dotted line represents the velocity just before the start of oxygen burning ($T_\mathrm{c}=9.45$).} 
    \label{fig:preheat-velocity}
\end{figure}

Figure \ref{fig:preheat-velocity} illustrates the velocity structure of $100M_\odot$ progenitors at two different central temperature values: $\log T_\mathrm{c}(\mathrm{K})=9.2$ (solid lines) and $\log T_\mathrm{c}(\mathrm{K})=9.45$ (dotted lines). It is important to note that the solid lines correspond to snapshots immediately preceding the triangle points, while the dotted lines correspond to snapshots as just before the square points in Figure \ref{fig:preheat-carbon}. Each color represents a different reaction rate ($-2\sigma$: blue, standard: green, $+2\sigma$: red).
At $\log T_\mathrm{c}(\mathrm{K})=9.2$ (solid lines), the velocity structure remains relatively consistent across all models, regardless of the reaction rate. The differences in infall speed are at most $\sim 0.2\times10^7$ cm s$^{-1}$. However, the preheating effect resulting from carbon combustion leads to noticeable variations in the infall velocity in low \cag rate environments, slowing it down by approximately $\sim 5\times10^7$ cm s$^{-1}$. Similarly, in high \cag rate environments, the infall progresses more rapidly, by $\sim 1\times10^7$ cm s$^{-1}$ (although not as fast as in the low (-2$\sigma$) and standard cases). Although this specific condition requires more detailed discussion, one possibility is that the enriched "preheating" may be making the star less compact. These findings indicate that carbon preheating contributes to the expansion of the star, rendering it "softer," and provide insights into the differential responses to subsequent explosive oxygen burning.

\begin{figure}
    \centering
    \includegraphics[width=0.48\textwidth]{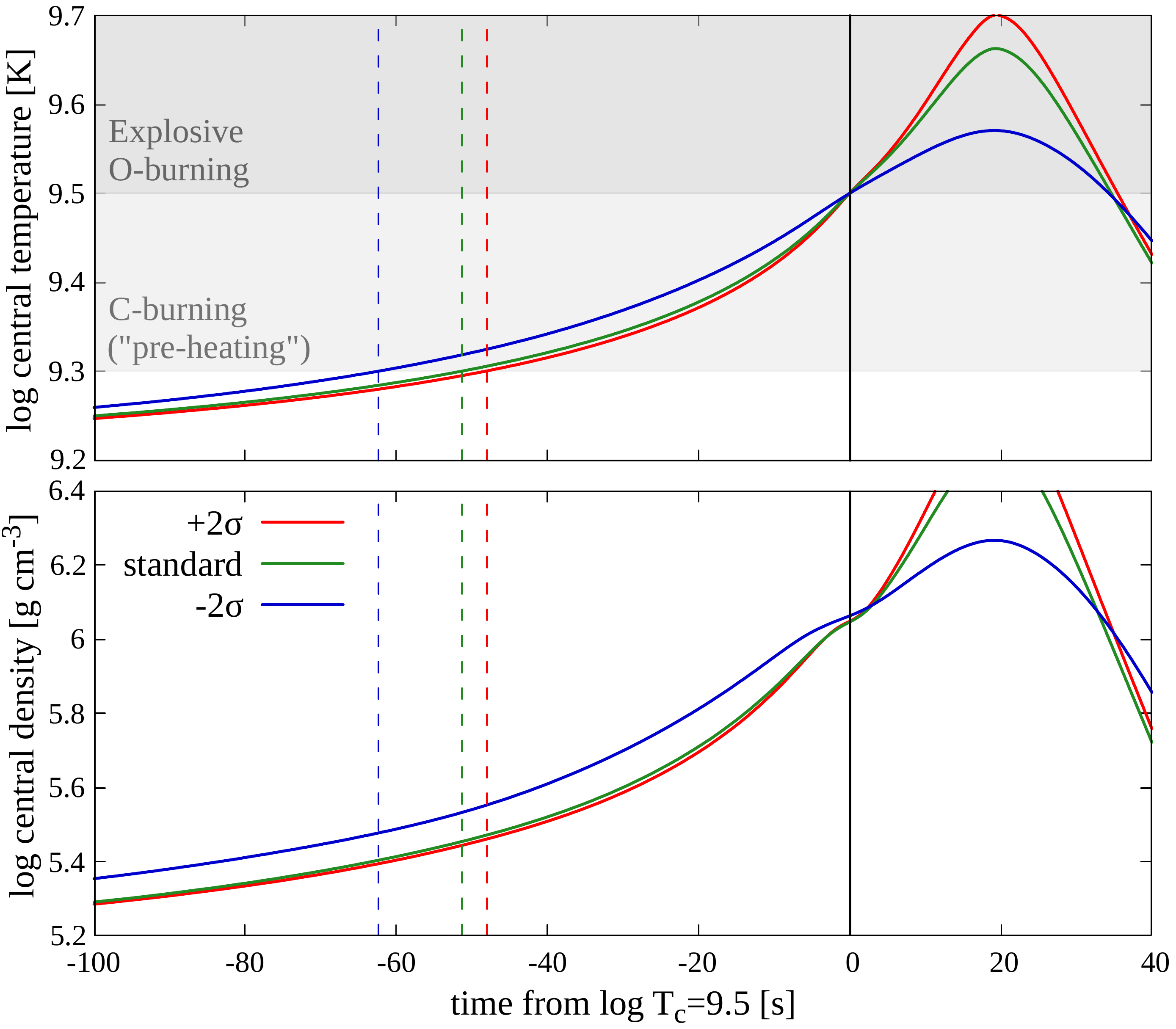}
    \caption{The time evolution of the central temperature (top panel) and central density (bottom panel) for models with an initial helium star mass of 100$M_{\odot}$. The dashed vertical lines represent the instances when $\log T_\mathrm{c}(\mathrm{K})=9.3$, indicating the initiation of carbon preheating.
} 
    \label{fig:preheat-evolution}
\end{figure}

In fact, the evolution during implosion is influenced by the expansion effect of carbon preheating. Figure \ref{fig:preheat-evolution} shows the time evolution of central temperature and central density in $100M_\odot$ progenitors. The dashed lines represent the time at $\log T_\mathrm{c}(\mathrm{K})=9.3$, marking the onset of carbon preheating. 
Focusing on these dashed lines, there are differences of ten seconds in the time from the start of preheating until reaching oxygen burning, depending on the reaction rate. Furthermore, it is observed that the lower the reaction rate, the slower the density increases in the preheating region.

These results suggest that stars with a significant amount of remaining carbon are capable of withstanding the dynamical compression with carbon preheating. While the evidence is not yet conclusive, we anticipate that this phenomenon contributes to the differences observed in the synthesized \nickel mass and explosion energy, as discussed in section \ref{sec:results}.

\section{The calculation about $^4$H\lowercase{e} $\rightarrow$ \lowercase{2n+2p} photodisintegration}\label{sec:appendix-calc}

In this appendix, we derivate Eq.(\ref{eq:photo}). Note that we use the following notations: "p" represents a proton, "n" represents a neutron, and "He" represents $^4$He.

Here, we assume that the nuclear formation and disintegration are in chemical equilibrium, described by the reaction:
\begin{align}
^4\mathrm{He}\rightleftharpoons 2\mathrm{p}+2\mathrm{n}-28.3\mathrm{MeV}.\label{eq:equilibrium}
\end{align}
Then, the abundances in nuclear equilibrium are given by the Saha's equation,
\begin{align}
\dfrac{n_\mathrm{p}^2n_\mathrm{n}^2}{n_\mathrm{He}}=\dfrac{g_\mathrm{p}^2g_\mathrm{g}^2}{g_\mathrm{He}}\left(\dfrac{2\pi k_\mathrm{B}T}{h^2}\right)^{9/2}\left(\dfrac{m_\mathrm{p}^2m_\mathrm{n}^2}{m_\mathrm{He}}\right)^{3/2}\exp\left(-\dfrac{Q}{k_\mathrm{B}T}\right),\label{eq:saha}
\end{align}
where $n_\mathrm{i}$ is the number density, $g_\mathrm{i}$ is the spin degree of freedom, and $m_\mathrm{i}$ is the mass for i particle, respectively. The number density is expressed by
\begin{align}
n_\mathrm{i}=\dfrac{\rho Y_\mathrm{i}}{m_\mathrm{i}},\label{eq:ntoT}
\end{align}
where $Y_\mathrm{i}$ is the number fraction and $\rho$ is the density.
From reaction \eqref{eq:equilibrium}, we note that $Q=28.3\mathrm{MeV}$. Also
\begin{align}
\dfrac{g_\mathrm{p}^2g_\mathrm{g}^2}{g_\mathrm{He}}=8 \:(\because g_\mathrm{p}=g_\mathrm{n}=g_\mathrm{He}=2),\label{eq:gcalc}
\end{align}
and 
\begin{align}
n_\mathrm{p}=n_\mathrm{n}\label{eq:nrelation}.
\end{align}
By assuming 
\begin{align}
m_\mathrm{p}\approx m_\mathrm{n}\approx \dfrac{m_\mathrm{He}}{4},
\label{eq:same}\end{align}
we get
\begin{align}
n_\mathrm{p}+n_\mathrm{n}+4n_\mathrm{He}=\dfrac{\rho}{m_\mathrm{p}}.
\end{align}
Combining these equations, we obtain
\begin{align}
Y_\mathrm{p}=&\dfrac{1}{2}(1-Y_\mathrm{He}),\label{eq:Yrel}
\end{align}
which gives
\begin{align}
n_{\rm p}=2\dfrac{1-Y_\mathrm{He}}{Y_\mathrm{He}}n_\mathrm{He}.
\end{align}
As a result, LHS of Eq. \eqref{eq:saha} is rewritten as
\begin{align}
\dfrac{n_\mathrm{p}^2n_\mathrm{n}^2}{n_\mathrm{He}}=&16\left(\dfrac{1-Y_\mathrm{He}}{Y_\mathrm{He}}\right)^4n_\mathrm{He}^3. 
\end{align}
Thus Eq. \eqref{eq:saha} reads
\begin{align}
{n_\mathrm{He}}=2^{-1/3}\left(\dfrac{1-Y_\mathrm{He}}{Y_\mathrm{He}}\right)^{-4/3}\left(\dfrac{2\pi k_\mathrm{B}T}{h^2}\right)^{3/2}\left(\dfrac{m_\mathrm{p}^2m_\mathrm{n}^2}{m_\mathrm{He}}\right)^{1/2}\exp\left(-\dfrac{Q}{3k_\mathrm{B}T}\right).
\end{align}
Substituting constants ($m_\mathrm{p}$, $m_\mathrm{n}$, $m_\mathrm{He}$, $h$ in cgs unit), we get
\begin{align}
\rho R(Y_\mathrm{He})=&7.9286\times10^{11}\left(\dfrac{k_\mathrm{B}T}{1\mathrm{MeV}}\right)^{3/2}\exp\left(-9.433\dfrac{1\mathrm{MeV}}{k_\mathrm{B}T}\right),\label{eq:rhoR}
\end{align}
where $R(Y_{\rm He})$ is presented by Eq. \eqref{eq:defR}.

Finally, taking the logarithm of Eq.(\ref{eq:rhoR}), we obtain
\begin{align}
    \log\left(\rho R(Y_\mathrm{He})\right)&=11.7974+\dfrac{3}{2}\log\left(\dfrac{k_\mathrm{B}T}{1\mathrm{MeV}}\right)-4.097\dfrac{1\mathrm{MeV}}{k_\mathrm{B}T},
\end{align}
which is Eq. \eqref{eq:photo}. 

\section{Comparison with the previous work}\label{sec:appendix-HW02}

In this appendix, we validate the reliability of our calculations by comparing them to previous studies that employed alternative calculation methods.

\begin{figure}
	\subfigure[The gained energy]{
		\includegraphics[width=0.48\textwidth]{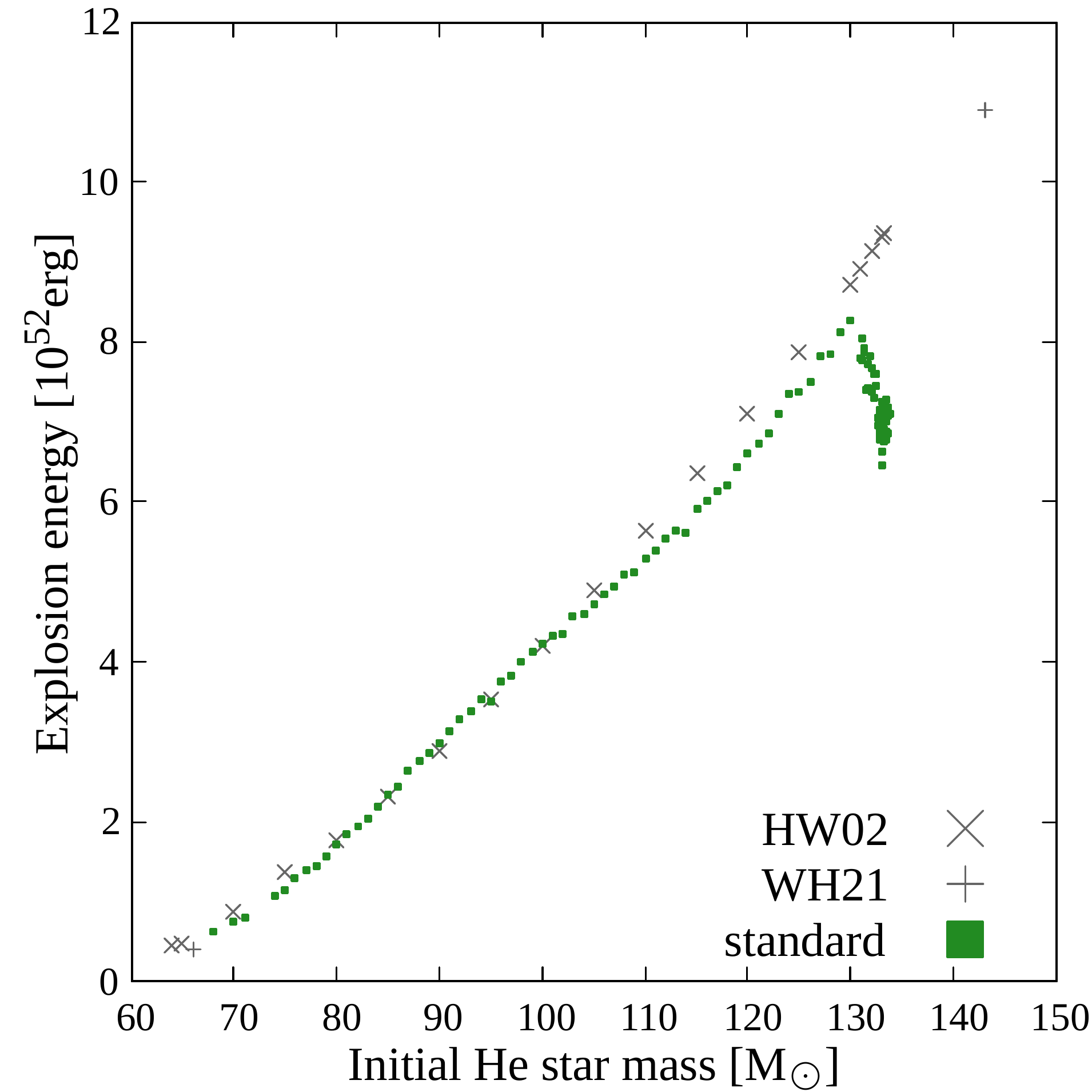}}\\ \label{pic:consistency-1}
	\subfigure[The amount of synthesized nickel]{
		\includegraphics[width=0.48\textwidth]{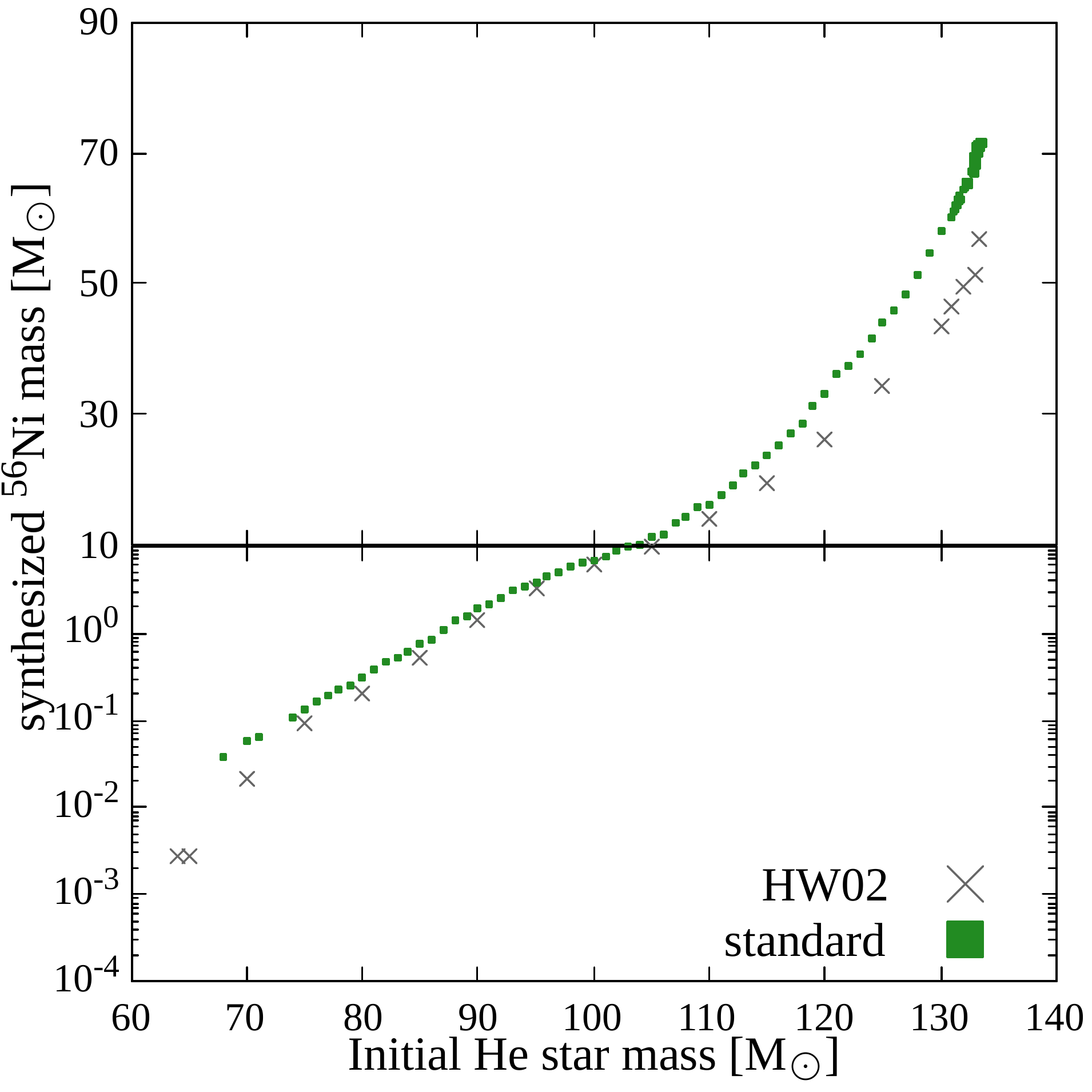}}\label{pic:consistency-2}
	\caption{The consistency with \citet{2002ApJ...567..532H}. The upper panel illustrates the relationship between the initial He core mass $M_\mathrm{init,He}$ and the energy gained. The lower panel shows the amount of synthesized nickel. The points in both panels represent our results (green points) and the results of \citet{2002ApJ...567..532H} as HW02 and \citet{2021ApJ...912L..31W} as WH21 (black points), indicating the consistency between the two studies.}
	\label{fig:consistency}
\end{figure}

Figure \ref{fig:consistency} illustrates the relationship between the initial He core mass $M_\mathrm{init,He}$ and (a) the final energy of the explosion $E_\mathrm{expl}$, and (b) the synthesized \nickel mass. These plots represent models with a standard \cag rate. We have included the results by \citet{2002ApJ...567..532H} for comparison. We confirm a positive correlation between the amount of synthesized \nickel and the explosion energy for the initial He core mass. Importantly, our results obtained without magnification of the \cag rate show reasonable consistency with previous studies.

Note that in Figure \ref{fig:consistency}(a), also in Figure \ref{fig:mass-energy}, we observe a drop in the explosion energy at the heavier end of the initial He core mass, which has not been reported in previous studies \citep[e.g.,][]{2002ApJ...567..532H}. It is possible that the explodable upper mass limit of PISNe, primarily governed by He photodisintegration, leads to the "freeze-out" of photodisintegrated elements from iron, resulting in a portion of the explosion energy being captured as rest mass energy. However, it is important to note that this is speculative, and we have not identified the exact physical cause of this trend.

\section{Data table}\label{sec:appendix-table}
We present values of the synthesized mass of \nickel and the explosion energy for all the models that explode as PISN resulting from this study in Table \ref{table:minus2} to \ref{table:plus2}. 

\clearpage
\begin{table}
\caption{: $-2\sigma$ series}
\label{table:minus2}
 \end{table}
 \begin{center}      
\begin{supertabular}{c|rr} 
\hline
    initial mass ($M_\odot$) & $E_\mathrm{expl}$ ($10^{51}\mathrm{erg}$) & $M_{^{56}\mathrm{Ni}}$ ($M_\odot$)\\
    \hline \hline 
91.0	&	12.511	&	0.007	\\
92.0	&	13.565	&	0.011	\\
93.0	&	14.491	&	0.012	\\
94.0	&	15.148	&	0.017	\\
95.0	&	16.188	&	0.023	\\
96.0	&	17.740	&	0.036	\\
97.0	&	17.905	&	0.037	\\
98.0	&	19.659	&	0.049	\\
99.0	&	20.991	&	0.063	\\
100.0	&	21.994	&	0.076	\\
101.0	&	22.973	&	0.088	\\
102.0	&	24.557	&	0.110	\\
103.0	&	26.543	&	0.153	\\
104.0	&	27.422	&	0.173	\\
105.0	&	29.083	&	0.215	\\
107.0	&	31.961	&	0.296	\\
108.0	&	33.276	&	0.367	\\
109.0	&	34.053	&	0.440	\\
110.0	&	36.177	&	0.586	\\
111.0	&	37.533	&	0.675	\\
112.0	&	39.465	&	0.796	\\
113.0	&	41.653	&	1.116	\\
115.0	&	44.569	&	1.495	\\
116.0	&	46.840	&	1.982	\\
117.0	&	48.306	&	2.196	\\
118.0	&	50.363	&	2.776	\\
119.0	&	51.553	&	2.993	\\
120.0	&	52.688	&	3.595	\\
121.0	&	54.672	&	4.206	\\
122.0	&	55.588	&	5.201	\\
123.0	&	57.190	&	5.306	\\
124.0	&	59.137	&	6.055	\\
125.0	&	60.993	&	7.022	\\
126.0	&	62.318	&	8.592	\\
127.0	&	64.802	&	8.785	\\
128.0	&	64.981	&	9.590	\\
129.0	&	66.932	&	10.486	\\
130.0	&	69.066	&	12.060	\\
131.0	&	70.850	&	12.810	\\
132.0	&	71.116	&	13.108	\\
133.0	&	73.365	&	14.787	\\
134.0	&	73.877	&	15.517	\\
135.0	&	75.343	&	16.799	\\
136.0	&	77.998	&	18.246	\\
137.0	&	78.989	&	19.712	\\
138.0	&	80.863	&	21.275	\\
139.0	&	81.876	&	22.789	\\
140.0	&	82.116	&	24.522	\\
141.0	&	84.378	&	26.383	\\
142.0	&	85.752	&	28.629	\\
143.0	&	87.495	&	30.692	\\
144.0	&	86.381	&	32.804	\\
145.0	&	90.063	&	34.538	\\
146.0	&	91.855	&	37.426	\\
147.0	&	93.466	&	39.716	\\
148.0	&	92.408	&	42.837	\\
149.0	&	97.966	&	46.389	\\
150.0	&	97.285	&	48.013	\\
151.0	&	100.351	&	50.357	\\
152.0	&	103.607	&	53.967	\\
153.0	&	105.196	&	57.709	\\
154.0	&	109.789	&	61.316	\\
155.0	&	110.906	&	62.822	\\
156.0	&	112.095	&	66.516	\\
157.0	&	118.701	&	70.121	\\
158.0	&	118.596	&	74.392	\\
159.0	&	104.138	&	77.990	\\
159.1	&	98.422	&	79.742	\\
159.2	&	96.523	&	80.150	\\
159.3	&	89.998	&	81.529	\\
159.4	&	93.981	&	80.780	\\
159.5	&	85.228	&	82.918	\\
159.6	&	89.653	&	82.020	\\
159.7	&	96.076	&	80.631	\\
159.8	&	84.733	&	83.524	\\
159.9	&	83.024	&	84.027	\\
160.0	&	88.001	&	82.443	\\
160.01	&	80.404	&	83.923	\\
160.02	&	75.921	&	83.394	\\
160.03	&	82.720	&	83.894	\\
160.04	&	89.225	&	82.663	\\
160.05	&	80.003	&	83.703	\\
160.06	&	80.491	&	83.555	\\
160.07	&	83.882	&	83.937	\\
160.08	&	77.789	&	83.212	\\
160.09	&	79.499	&	83.897	\\
160.1	&	82.440	&	83.847	\\
160.11	&	80.656	&	83.899	\\
160.12	&	86.731	&	83.051	\\
160.13	&	73.554	&	83.213	\\
160.14	&	89.761	&	82.450	\\
160.15	&	77.074	&	83.725	\\
160.16	&	79.719	&	84.010	\\
160.22	&	76.173	&	83.526	\\
160.23	&	77.977	&	83.741	\\
160.24	&	76.391	&	83.373	\\
160.25	&	76.217	&	83.223	\\
160.26	&	76.932	&	83.321	\\
160.28	&	76.281	&	83.346	\\
160.29	&	77.428	&	83.305	\\
160.31	&	76.168	&	83.351	\\
160.32	&	80.917	&	83.710	\\
160.33	&	79.453	&	83.396	\\
160.49	&	80.745	&	83.559	\\
160.52	&	76.525	&	83.508	\\
    \hline
\end{supertabular}
\end{center}

\newpage
\begin{table}

\caption{: $-1\sigma$ series}
\label{table:minus1}
 \end{table}
 \begin{center}
\begin{supertabular}{c|rr} \hline
    initial mass ($M_\odot$) & $E_\mathrm{expl}$ ($10^{51}\mathrm{erg}$) & $M_{^{56}\mathrm{Ni}}$ ($M_\odot$)\\
    \hline \hline 
74.0	&	7.170	&	0.015	\\
79.0	&	9.399	&	0.024	\\
80.0	&	9.714	&	0.026	\\
81.0	&	11.255	&	0.040	\\
82.0	&	12.849	&	0.055	\\
83.0	&	13.463	&	0.076	\\
84.0	&	14.358	&	0.086	\\
85.0	&	15.355	&	0.101	\\
86.0	&	16.787	&	0.121	\\
87.0	&	17.663	&	0.150	\\
88.0	&	18.500	&	0.168	\\
89.0	&	19.957	&	0.182	\\
90.0	&	21.099	&	0.213	\\
91.0	&	21.642	&	0.260	\\
92.0	&	23.826	&	0.335	\\
93.0	&	24.461	&	0.402	\\
94.0	&	25.751	&	0.412	\\
95.0	&	26.964	&	0.523	\\
96.0	&	29.625	&	0.726	\\
97.0	&	31.303	&	0.869	\\
98.0	&	32.449	&	0.981	\\
99.0	&	34.073	&	1.206	\\
100.0	&	35.767	&	1.701	\\
101.0	&	36.711	&	1.744	\\
102.0	&	38.357	&	2.178	\\
103.0	&	39.083	&	2.522	\\
104.0	&	40.338	&	2.652	\\
105.0	&	42.147	&	3.292	\\
106.0	&	43.150	&	3.984	\\
107.0	&	44.906	&	4.233	\\
108.0	&	46.386	&	5.052	\\
109.0	&	47.471	&	5.451	\\
110.0	&	49.144	&	6.656	\\
111.0	&	50.694	&	7.341	\\
112.0	&	51.694	&	8.260	\\
113.0	&	53.226	&	9.224	\\
114.0	&	54.030	&	9.820	\\
115.0	&	55.743	&	10.569	\\
116.0	&	57.082	&	11.669	\\
117.0	&	58.430	&	12.263	\\
118.0	&	59.269	&	13.408	\\
119.0	&	60.361	&	14.256	\\
120.0	&	61.457	&	15.980	\\
121.0	&	62.898	&	16.536	\\
122.0	&	63.296	&	18.233	\\
123.0	&	64.621	&	19.771	\\
124.0	&	65.938	&	21.064	\\
125.0	&	67.707	&	22.702	\\
126.0	&	68.622	&	24.241	\\
127.0	&	69.045	&	26.123	\\
128.0	&	70.868	&	27.851	\\
129.0	&	71.624	&	30.474	\\
130.0	&	74.146	&	32.511	\\
131.0	&	76.305	&	34.457	\\
132.0	&	76.731	&	37.393	\\
133.0	&	79.800	&	39.175	\\
134.0	&	81.016	&	41.012	\\
135.0	&	84.518	&	43.547	\\
136.0	&	83.404	&	46.625	\\
137.0	&	89.058	&	49.512	\\
138.0	&	89.519	&	52.361	\\
139.0	&	91.803	&	55.047	\\
140.0	&	90.872	&	58.948	\\
141.0	&	94.032	&	60.384	\\
142.0	&	97.760	&	64.096	\\
143.0	&	90.491	&	69.568	\\
144.0	&	86.285	&	71.598	\\
144.01	&	85.633	&	71.902	\\
144.02	&	83.705	&	72.380	\\
144.03	&	83.856	&	72.417	\\
144.04	&	81.452	&	72.970	\\
144.05	&	84.354	&	72.144	\\
144.06	&	83.028	&	72.477	\\
144.07	&	81.435	&	72.296	\\
144.08	&	86.281	&	71.683	\\
144.09	&	77.677	&	73.805	\\
144.1	&	83.377	&	72.524	\\
144.11	&	85.262	&	71.912	\\
144.12	&	83.812	&	72.445	\\
144.13	&	84.712	&	72.191	\\
144.14	&	78.819	&	73.635	\\
144.15	&	82.745	&	72.670	\\
144.16	&	81.625	&	72.870	\\
144.17	&	80.822	&	73.170	\\
144.18	&	83.814	&	72.396	\\
144.19	&	84.197	&	72.469	\\
144.2	&	82.223	&	72.834	\\
144.21	&	76.990	&	74.119	\\
144.22	&	72.112	&	76.712	\\
144.23	&	81.681	&	72.950	\\
144.24	&	83.911	&	72.346	\\
144.25	&	82.540	&	72.780	\\
144.26	&	82.702	&	72.858	\\
144.27	&	72.887	&	75.738	\\
144.28	&	72.693	&	75.593	\\
144.29	&	68.573	&	76.750	\\
144.3	&	80.757	&	73.438	\\
144.31	&	81.063	&	73.303	\\
144.32	&	83.499	&	72.629	\\
144.33	&	81.958	&	73.052	\\
144.34	&	77.951	&	74.020	\\
144.35	&	76.581	&	74.347	\\
144.36	&	81.908	&	73.198	\\
144.37	&	80.035	&	73.670	\\
144.38	&	76.972	&	74.246	\\
144.39	&	73.778	&	74.717	\\
144.4	&	79.943	&	73.615	\\
144.41	&	82.622	&	72.953	\\
144.42	&	77.958	&	73.928	\\
144.43	&	77.921	&	74.111	\\
144.44	&	75.815	&	74.624	\\
144.45	&	71.716	&	76.499	\\
144.46	&	74.036	&	75.179	\\
144.47	&	76.394	&	74.440	\\
144.48	&	73.622	&	75.416	\\
144.49	&	68.581	&	76.881	\\
144.5	&	78.349	&	74.113	\\
144.51	&	72.385	&	76.474	\\
144.52	&	72.789	&	76.120	\\
144.53	&	77.174	&	74.362	\\
144.54	&	72.482	&	76.219	\\
144.55	&	76.713	&	74.614	\\
144.56	&	72.192	&	76.823	\\
144.57	&	75.079	&	75.104	\\
144.58	&	76.267	&	74.690	\\
144.59	&	71.971	&	76.787	\\
144.6	&	74.448	&	75.172	\\
144.61	&	72.558	&	76.010	\\
144.62	&	74.702	&	74.969	\\
144.63	&	73.984	&	75.579	\\
144.64	&	75.595	&	74.834	\\
144.65	&	76.052	&	74.633	\\
144.67	&	74.559	&	75.207	\\
144.68	&	72.878	&	75.977	\\
144.69	&	73.049	&	76.916	\\
144.7	&	72.216	&	76.616	\\
144.71	&	72.662	&	76.115	\\
144.72	&	73.863	&	75.619	\\
144.73	&	72.261	&	75.719	\\
144.74	&	72.217	&	76.679	\\
144.75	&	72.552	&	76.270	\\
144.76	&	76.809	&	74.542	\\
144.77	&	72.186	&	76.926	\\
144.78	&	75.469	&	75.064	\\
144.79	&	73.237	&	75.825	\\
144.8	&	71.954	&	76.714	\\
144.81	&	71.624	&	76.426	\\
144.85	&	72.569	&	76.883	\\
144.86	&	69.570	&	76.470	\\
144.87	&	73.009	&	76.410	\\
144.88	&	72.550	&	76.718	\\
144.89	&	71.126	&	76.455	\\
144.9	&	71.981	&	76.782	\\
144.91	&	72.443	&	76.965	\\
144.92	&	73.830	&	75.848	\\
144.93	&	71.598	&	76.420	\\
144.95	&	71.435	&	76.474	\\
144.96	&	71.441	&	76.504	\\
144.97	&	72.527	&	76.904	\\
144.98	&	72.374	&	76.993	\\
144.99	&	71.910	&	76.485	\\
145.0	&	72.699	&	76.524	\\
145.01	&	71.698	&	76.924	\\
145.02	&	72.389	&	77.083	\\
145.03	&	72.633	&	76.705	\\
145.04	&	71.796	&	76.561	\\
145.06	&	70.201	&	76.961	\\
145.07	&	72.801	&	76.874	\\
145.12	&	70.323	&	76.848	\\
145.14	&	72.277	&	76.943	\\
    \hline
\end{supertabular}
\end{center}

\newpage
\begin{table}
\caption{: Standard series}
\label{table:standard}
 \end{table}
\begin{center}    
\begin{supertabular}{c|rr} \hline
    initial mass ($M_\odot$) & $E_\mathrm{expl}$ ($10^{51}\mathrm{erg}$) & $M_{^{56}\mathrm{Ni}}$ ($M_\odot$)\\
    \hline \hline 
68.0	&	6.240	&	0.038	\\
70.0	&	7.495	&	0.058	\\
71.0	&	7.997	&	0.063	\\
74.0	&	10.664	&	0.111	\\
75.0	&	11.567	&	0.131	\\
76.0	&	13.099	&	0.167	\\
77.0	&	14.117	&	0.194	\\
78.0	&	14.444	&	0.222	\\
79.0	&	15.796	&	0.248	\\
80.0	&	17.235	&	0.314	\\
81.0	&	18.502	&	0.377	\\
82.0	&	19.549	&	0.459	\\
83.0	&	20.475	&	0.525	\\
84.0	&	21.892	&	0.610	\\
85.0	&	23.527	&	0.744	\\
86.0	&	24.359	&	0.823	\\
87.0	&	26.418	&	1.077	\\
88.0	&	27.571	&	1.383	\\
89.0	&	28.677	&	1.599	\\
90.0	&	29.829	&	1.915	\\
91.0	&	31.295	&	2.191	\\
92.0	&	32.732	&	2.555	\\
93.0	&	33.817	&	3.070	\\
94.0	&	35.235	&	3.407	\\
95.0	&	35.086	&	3.770	\\
96.0	&	37.565	&	4.494	\\
97.0	&	38.265	&	4.848	\\
98.0	&	39.988	&	5.686	\\
99.0	&	41.207	&	6.418	\\
100.0	&	42.164	&	6.870	\\
101.0	&	43.241	&	7.677	\\
102.0	&	43.617	&	8.629	\\
103.0	&	45.626	&	9.608	\\
104.0	&	46.025	&	10.053	\\
105.0	&	47.227	&	11.325	\\
106.0	&	48.530	&	11.784	\\
107.0	&	49.430	&	13.447	\\
108.0	&	51.047	&	14.256	\\
109.0	&	51.248	&	15.787	\\
110.0	&	52.903	&	16.162	\\
111.0	&	53.972	&	17.695	\\
112.0	&	55.364	&	19.234	\\
113.0	&	56.346	&	20.971	\\
114.0	&	56.191	&	22.286	\\
115.0	&	59.153	&	23.929	\\
116.0	&	60.197	&	25.445	\\
117.0	&	61.372	&	27.168	\\
118.0	&	62.123	&	28.597	\\
119.0	&	64.322	&	31.449	\\
120.0	&	66.102	&	33.266	\\
121.0	&	67.358	&	36.060	\\
122.0	&	68.572	&	37.484	\\
123.0	&	70.939	&	39.380	\\
124.0	&	73.493	&	41.802	\\
125.0	&	73.676	&	44.179	\\
126.0	&	75.105	&	46.058	\\
127.0	&	78.144	&	48.446	\\
128.0	&	78.363	&	51.229	\\
129.0	&	81.199	&	54.849	\\
130.0	&	82.802	&	57.982	\\
131.0	&	77.870	&	60.243	\\
131.1	&	80.579	&	60.988	\\
131.2	&	77.673	&	61.361	\\
131.3	&	78.743	&	62.084	\\
131.4	&	79.245	&	62.103	\\
131.5	&	74.003	&	63.032	\\
131.6	&	77.129	&	62.596	\\
131.7	&	74.163	&	63.534	\\
131.8	&	78.273	&	62.984	\\
132.0	&	76.808	&	64.319	\\
132.1	&	73.683	&	65.472	\\
132.2	&	76.020	&	64.682	\\
132.3	&	73.012	&	65.362	\\
132.4	&	75.939	&	65.109	\\
132.5	&	74.425	&	65.758	\\
132.6	&	69.627	&	67.037	\\
132.7	&	70.423	&	67.208	\\
132.8	&	71.065	&	67.150	\\
132.81	&	69.740	&	67.747	\\
132.82	&	70.605	&	67.267	\\
132.83	&	71.493	&	66.884	\\
132.84	&	68.489	&	68.637	\\
132.85	&	68.232	&	69.051	\\
132.87	&	69.666	&	67.626	\\
132.88	&	68.688	&	68.494	\\
132.89	&	69.271	&	67.459	\\
132.9	&	67.826	&	69.541	\\
132.91	&	69.079	&	68.331	\\
132.92	&	68.757	&	68.276	\\
132.93	&	69.012	&	68.783	\\
132.94	&	69.801	&	67.671	\\
132.95	&	70.071	&	67.675	\\
132.96	&	71.615	&	66.949	\\
132.97	&	69.576	&	67.895	\\
132.98	&	66.422	&	69.994	\\
132.99	&	70.380	&	67.530	\\
133.0	&	68.945	&	69.203	\\
133.01	&	67.950	&	69.289	\\
133.02	&	68.233	&	69.501	\\
133.03	&	67.882	&	69.616	\\
133.04	&	72.622	&	71.147	\\
133.05	&	68.008	&	69.476	\\
133.07	&	67.887	&	70.058	\\
133.08	&	68.565	&	69.092	\\
133.09	&	69.631	&	67.935	\\
133.11	&	67.741	&	69.799	\\
133.12	&	68.837	&	69.007	\\
133.13	&	68.415	&	69.356	\\
133.14	&	64.707	&	70.551	\\
133.15	&	68.836	&	69.153	\\
133.16	&	68.703	&	69.390	\\
133.17	&	67.640	&	69.238	\\
133.18	&	68.021	&	70.517	\\
133.19	&	70.346	&	71.512	\\
133.2	&	70.178	&	71.449	\\
133.21	&	68.927	&	70.570	\\
133.22	&	67.839	&	69.816	\\
133.23	&	68.270	&	69.708	\\
133.24	&	68.228	&	69.914	\\
133.25	&	70.338	&	71.353	\\
133.26	&	68.241	&	69.842	\\
133.27	&	71.259	&	71.541	\\
133.28	&	70.653	&	71.444	\\
133.29	&	70.884	&	71.476	\\
133.3	&	69.151	&	70.727	\\
133.31	&	68.132	&	70.425	\\
133.33	&	70.474	&	71.379	\\
133.34	&	70.104	&	70.973	\\
133.35	&	67.975	&	70.174	\\
133.36	&	68.140	&	69.969	\\
133.37	&	67.801	&	70.195	\\
133.39	&	71.438	&	71.221	\\
133.4	&	72.776	&	71.554	\\
133.41	&	68.319	&	70.620	\\
133.42	&	71.298	&	71.656	\\
133.43	&	71.078	&	71.460	\\
133.46	&	68.388	&	70.689	\\
133.47	&	71.315	&	71.060	\\
133.48	&	71.263	&	71.274	\\
133.49	&	68.875	&	71.311	\\
133.52	&	71.036	&	71.663	\\
133.54	&	71.741	&	71.410	\\
133.55	&	71.011	&	71.189	\\
133.56	&	71.902	&	71.549	\\
133.57	&	71.146	&	71.639	\\
133.58	&	70.759	&	71.569	\\
133.6	&	71.222	&	71.476	\\
133.61	&	71.486	&	71.657	\\
133.62	&	68.564	&	71.417	\\
133.63	&	71.531	&	71.509	\\
133.72	&	70.959	&	71.471	\\
    \hline
\end{supertabular}
\end{center}

\newpage
\begin{table}
\caption{: $+1\sigma$ series}
\label{table:plus1}
 \end{table}
\begin{center}    
\begin{supertabular}{c|rr} \hline
    initial mass ($M_\odot$) & $E_\mathrm{expl}$ ($10^{51}\mathrm{erg}$) & $M_{^{56}\mathrm{Ni}}$ ($M_\odot$)\\
    \hline \hline 
63.0	&	4.909	&	0.057	\\
64.0	&	5.530	&	0.068	\\
65.0	&	6.386	&	0.093	\\
66.0	&	7.188	&	0.112	\\
67.0	&	7.903	&	0.127	\\
68.0	&	8.859	&	0.161	\\
69.0	&	9.796	&	0.179	\\
70.0	&	10.553	&	0.208	\\
71.0	&	11.627	&	0.249	\\
74.0	&	14.984	&	0.396	\\
75.0	&	16.287	&	0.464	\\
76.0	&	17.677	&	0.547	\\
77.0	&	18.608	&	0.619	\\
78.0	&	19.847	&	0.809	\\
79.0	&	20.839	&	0.967	\\
80.0	&	22.200	&	1.143	\\
81.0	&	23.768	&	1.362	\\
82.0	&	24.787	&	1.576	\\
83.0	&	25.249	&	1.815	\\
84.0	&	27.232	&	2.150	\\
85.0	&	27.931	&	2.402	\\
86.0	&	29.258	&	2.862	\\
87.0	&	30.707	&	3.375	\\
88.0	&	31.514	&	3.870	\\
89.0	&	33.227	&	4.456	\\
90.0	&	33.615	&	4.976	\\
91.0	&	34.728	&	5.304	\\
92.0	&	36.046	&	6.034	\\
93.0	&	37.144	&	6.593	\\
94.0	&	37.376	&	7.351	\\
95.0	&	38.535	&	8.127	\\
96.0	&	39.742	&	8.986	\\
97.0	&	40.792	&	9.895	\\
98.0	&	41.832	&	10.773	\\
99.0	&	41.310	&	11.813	\\
100.0	&	43.497	&	12.852	\\
101.0	&	45.750	&	13.571	\\
102.0	&	45.761	&	14.465	\\
103.0	&	47.208	&	15.755	\\
104.0	&	48.032	&	17.312	\\
105.0	&	49.331	&	18.219	\\
106.0	&	51.329	&	20.039	\\
107.0	&	51.282	&	20.848	\\
108.0	&	52.897	&	22.548	\\
109.0	&	53.983	&	23.793	\\
110.0	&	55.974	&	25.831	\\
111.0	&	57.185	&	27.934	\\
112.0	&	57.329	&	29.424	\\
113.0	&	58.524	&	30.888	\\
114.0	&	59.956	&	32.975	\\
115.0	&	62.742	&	34.845	\\
116.0	&	64.067	&	37.750	\\
117.0	&	65.201	&	39.029	\\
118.0	&	67.094	&	41.626	\\
119.0	&	68.388	&	43.900	\\
120.0	&	69.560	&	46.244	\\
121.0	&	71.508	&	49.104	\\
122.0	&	72.343	&	51.352	\\
123.0	&	73.971	&	54.133	\\
124.0	&	71.545	&	57.078	\\
124.1	&	68.304	&	58.381	\\
124.2	&	71.196	&	57.739	\\
124.3	&	70.441	&	59.580	\\
124.4	&	67.706	&	59.025	\\
124.5	&	71.687	&	58.751	\\
124.6	&	68.789	&	60.575	\\
124.7	&	66.657	&	58.898	\\
124.8	&	71.154	&	59.585	\\
124.9	&	69.422	&	60.663	\\
125.0	&	67.480	&	61.562	\\
125.1	&	64.969	&	61.626	\\
125.2	&	66.857	&	61.729	\\
125.3	&	67.251	&	62.142	\\
125.4	&	66.495	&	62.512	\\
125.5	&	65.041	&	63.224	\\
125.6	&	63.482	&	64.032	\\
125.7	&	64.067	&	63.826	\\
125.8	&	63.807	&	64.860	\\
125.9	&	63.441	&	64.361	\\
126.0	&	64.279	&	64.180	\\
126.1	&	63.606	&	65.612	\\
126.2	&	62.565	&	66.111	\\
126.3	&	64.052	&	65.473	\\
126.4	&	63.499	&	66.653	\\
126.5	&	65.599	&	67.185	\\
126.6	&	68.805	&	68.311	\\
126.7	&	67.741	&	67.347	\\
126.8	&	70.013	&	68.389	\\
126.82	&	69.987	&	68.374	\\
126.84	&	69.164	&	68.457	\\
126.94	&	71.753	&	68.328	\\
127.01	&	70.860	&	68.431	\\
    \hline
\end{supertabular}
\end{center}

\newpage
\begin{table}
\caption{: $+2\sigma$ series}
\label{table:plus2}
 \end{table}
\begin{center}    
\begin{supertabular}{c|rr} \hline
    initial mass ($M_\odot$) & $E_\mathrm{expl}$ ($10^{51}\mathrm{erg}$) & $M_{^{56}\mathrm{Ni}}$ ($M_\odot$)\\
    \hline \hline 
62.0	&	5.168	&	0.094	\\
63.0	&	5.601	&	0.107	\\
64.0	&	6.771	&	0.146	\\
65.0	&	7.318	&	0.160	\\
66.0	&	8.282	&	0.200	\\
67.0	&	9.115	&	0.228	\\
68.0	&	10.090	&	0.278	\\
69.0	&	10.873	&	0.314	\\
70.0	&	11.924	&	0.372	\\
71.0	&	13.259	&	0.420	\\
74.0	&	16.293	&	0.671	\\
75.0	&	17.978	&	0.812	\\
76.0	&	19.010	&	0.925	\\
77.0	&	20.149	&	1.081	\\
78.0	&	21.031	&	1.256	\\
79.0	&	22.249	&	1.544	\\
80.0	&	23.523	&	1.742	\\
81.0	&	24.837	&	1.982	\\
82.0	&	25.629	&	2.409	\\
83.0	&	26.943	&	2.804	\\
84.0	&	28.474	&	3.173	\\
85.0	&	29.029	&	3.563	\\
86.0	&	30.532	&	4.180	\\
87.0	&	31.487	&	4.634	\\
88.0	&	32.323	&	5.263	\\
89.0	&	33.357	&	5.956	\\
90.0	&	34.177	&	6.609	\\
91.0	&	35.029	&	7.210	\\
92.0	&	35.650	&	7.794	\\
93.0	&	37.320	&	8.841	\\
94.0	&	38.419	&	9.638	\\
95.0	&	39.351	&	10.456	\\
96.0	&	40.946	&	11.014	\\
97.0	&	41.039	&	12.424	\\
98.0	&	42.794	&	13.369	\\
99.0	&	43.395	&	14.086	\\
100.0	&	44.096	&	15.392	\\
101.0	&	45.714	&	16.049	\\
102.0	&	46.174	&	17.791	\\
103.0	&	47.341	&	19.134	\\
104.0	&	48.465	&	20.343	\\
105.0	&	50.321	&	21.532	\\
106.0	&	51.305	&	23.892	\\
107.0	&	52.732	&	24.599	\\
108.0	&	53.743	&	27.496	\\
109.0	&	54.998	&	28.276	\\
110.0	&	55.483	&	30.123	\\
111.0	&	57.199	&	32.132	\\
112.0	&	58.933	&	33.544	\\
113.0	&	59.932	&	35.889	\\
114.0	&	61.076	&	38.316	\\
115.0	&	63.123	&	40.034	\\
116.0	&	63.940	&	42.771	\\
117.0	&	66.161	&	44.690	\\
118.0	&	68.065	&	47.874	\\
119.0	&	68.997	&	50.647	\\
120.0	&	70.208	&	53.058	\\
121.0	&	70.360	&	55.208	\\
122.0	&	64.158	&	58.613	\\
122.1	&	68.107	&	58.602	\\
122.2	&	66.929	&	59.605	\\
122.3	&	65.394	&	60.342	\\
122.4	&	64.910	&	60.743	\\
122.5	&	64.897	&	60.986	\\
122.6	&	63.419	&	61.801	\\
122.7	&	65.228	&	60.991	\\
122.8	&	64.114	&	61.564	\\
122.9	&	65.710	&	60.871	\\
123.0	&	61.331	&	62.886	\\
123.1	&	63.446	&	62.239	\\
123.2	&	62.255	&	62.935	\\
123.3	&	61.668	&	63.326	\\
123.4	&	60.687	&	65.230	\\
123.5	&	61.747	&	64.110	\\
123.51	&	61.447	&	64.023	\\
123.52	&	62.790	&	65.918	\\
123.53	&	61.549	&	64.787	\\
123.54	&	61.548	&	65.766	\\
123.55	&	60.147	&	65.254	\\
123.56	&	62.826	&	65.865	\\
123.57	&	67.677	&	67.828	\\
123.58	&	63.448	&	66.479	\\
123.59	&	61.153	&	65.534	\\
123.6	&	61.302	&	64.261	\\
123.61	&	65.898	&	66.804	\\
123.62	&	62.101	&	65.720	\\
123.63	&	61.594	&	65.054	\\
123.64	&	61.831	&	64.129	\\
123.65	&	67.152	&	67.388	\\
123.66	&	66.822	&	66.942	\\
123.67	&	67.618	&	67.739	\\
123.68	&	67.422	&	67.129	\\
123.69	&	62.012	&	65.757	\\
123.7	&	61.935	&	64.768	\\
123.71	&	65.099	&	66.553	\\
123.72	&	67.810	&	67.841	\\
123.73	&	67.462	&	67.800	\\
123.74	&	67.587	&	67.395	\\
123.75	&	67.661	&	67.750	\\
123.76	&	67.374	&	67.832	\\
123.77	&	65.010	&	66.515	\\
123.78	&	66.153	&	66.885	\\
123.79	&	66.355	&	66.812	\\
123.8	&	66.547	&	67.720	\\
123.81	&	65.491	&	66.499	\\
123.82	&	65.223	&	66.380	\\
123.83	&	68.037	&	67.732	\\
123.84	&	63.912	&	66.333	\\
123.86	&	68.657	&	67.671	\\
123.87	&	68.888	&	67.649	\\
123.89	&	61.794	&	65.516	\\
123.9	&	66.043	&	66.694	\\
123.92	&	66.373	&	67.842	\\
123.93	&	66.742	&	67.330	\\
123.94	&	66.909	&	67.110	\\
123.96	&	69.205	&	67.526	\\
123.97	&	67.497	&	67.685	\\
123.98	&	68.122	&	67.855	\\
123.99	&	68.304	&	67.889	\\
124.01	&	67.928	&	67.724	\\
124.02	&	67.844	&	67.797	\\
124.04	&	66.779	&	67.160	\\
124.06	&	68.770	&	67.596	\\
124.08	&	67.790	&	67.015	\\
124.09	&	67.828	&	67.657	\\
124.1	&	66.976	&	67.763	\\
124.15	&	70.237	&	67.604	\\
    \hline
\end{supertabular}
\end{center}


\bsp	
\label{lastpage}
\end{document}